\documentclass[printer]{aa}
\usepackage{graphicx}
\usepackage{txfonts}
\usepackage{natbib}
\usepackage{multirow}
\usepackage{slashbox}

\addtocounter{secnumdepth}{2}

\begin{document}

\title{Self-coherent camera as a focal plane\\wavefront sensor: simulations.}

\author{Rapha\"{e}l~Galicher\inst{1}\fnmsep\inst{2}, Pierre~Baudoz\inst{1}\fnmsep\inst{2}, G\'{e}rard~Rousset\inst{1}\fnmsep\inst{2}, Julien~Totems\inst{2}\fnmsep\inst{3}, Marion~Mas\inst{1}\fnmsep\inst{2}}
\institute{LESIA, Observatoire de Paris, CNRS, University Pierre et Marie Curie Paris 6 and University Denis Diderot Paris 7. 5, place Jules~Janssen, 92195 Meudon, France.\\
  \email{raphael.galicher@obspm.fr, pierre.baudoz@obspm.fr,gerard.rousset@obspm.fr}
 \and
 Groupement d'Int\'{e}r\^{e}t Scientifique Partenariat Haute R\'{e}solution Angulaire Sol Espace~(PHASE) between ONERA, Observatoire de Paris, CNRS and University Denis Diderot Paris 7
 \and
 Onera / DOTA - Chemin de la Huni\`ere - 91761 Palaiseau Cedex\\
\email{julien.totems@onera.fr}}

\abstract
    {Direct detection of exoplanets requires high dynamic range imaging. Coronagraphs could be the solution, but their performance in space is limited by wavefront errors~(manufacturing errors on optics, temperature variations, etc.), which create quasi-static stellar speckles in the final image.}
    {Several solutions have been suggested for tackling this speckle noise. Differential imaging techniques substract a reference image to the coronagraphic residue in a post-processing imaging. Other techniques attempt to actively correct wavefront errors using a deformable mirror. In that case, wavefront aberrations have to be measured in the science image to extremely high accuracy.}
    {We propose the self-coherent camera sequentially used as a focal-plane wavefront sensor for active correction and differential imaging. For both uses, stellar speckles are spatially encoded in the science image so that differential aberrations are strongly minimized. The encoding is based on the principle of light incoherence between the hosting star and its environment.}
    {In this paper, we first discuss one intrinsic limitation of deformable mirrors. Then, several parameters of the self-coherent camera are studied in detail. We also propose an easy and robust design to associate the self-coherent camera with a coronagraph that uses a Lyot stop. Finally, we discuss the case of the association with a four-quadrant phase mask and numerically demonstrate that such a device enables detection of Earth-like planets under realistic conditions.}
    {The parametric study of the technique lets us believe it can be implemented quite easily in future instruments dedicated to direct imaging of exoplanets.}
    
    \date{Received ; accepted }
    
    \titlerunning{Parametric study of SCC}
    \authorrunning{R.~Galicher et al.}
    
    \maketitle
    
    \section{Introduction}
    Exoplanets are typically~$10^7$ to~$10^{10}$ fainter than their host and are often located within a fraction of an arcsecond from their star. Numerous coronagraphs have been proposed to reduce the overwhelming light of a star to obtain a direct imaging of extrasolar planets~\citep{Rouan00,Mawet05,Guyon05}. Several of them provide observations~\citep{SchneiderG98,Boccaletti04}. But performance is limited by wavefront errors in the upstream beam for all these coronagraphs and the final focal plane image shows stellar speckles. The effect of most of these aberrations can be corrected by adaptive optics~(AO) or eXtreme AO~\citep[XAO,][]{Verinaud08} but the uncorrected part generates quasi-static speckles, which limit the image contrast\,\citep{Cavarroc06,macintosh05}. To reduce this speckle noise, differential imaging techniques attempt to subtract a reference image of the stellar speckles from the science image~(star plus companion).

Several ways are used to measure that reference: spectral characteristics~\citep{racine99,marois00,marois04}, polarization states~\citep{Baba03,Stam04}, differential rotation in image~\citep{SchneiderG98b,marois06}, or incoherence between stellar and companion lights~\citep{Guyon04}. The self-coherent camera with which we work uses the last property~\citep{Baudoz06,Galicher07}. But before using one of these {\it a posteriori} techniques, we may actively correct quasi-static wavefront errors so that a first speckle reduction is achieved and differential imaging techniques have less to do.  Because of the low level of aberrations that must be achieved~(a few nanometers), the wavefront sensor of that loop has to measure for phase and amplitude errors in the final science image to avoid differential errors introduced by classical wavefront sensor~\citep[Shack-Hartmann for example,][]{shack71}. \citet{Codona04} suggest using the incoherence between companion and stellar lights and use a modified Mach-Zender interferometer to encoded the stellar speckles.

The instrument we propose, the self coherent camera~(SCC), is based on the same property and uses Fizeau interferences. We insist on how~SCC can be used both as a wavefront sensor for active correction~(called step~A in this paper) and as a differential imaging technique~(step~B). In~\citet{Galicher08}, we numerically demonstrated that, applying step~B after step~A, a self-coherent camera associated with a~$32$x$32$ deformable mirror and a perfect coronagraph detects earths~(contrast of~$2\,10^{-10}$) from space in a few hours under realistic assumptions~(zodiacal light, photon noise, read-out noise, phase errors of~$20$\,nm rms, $20\%$~bandwidth and $1\%$~effective bandwidth,~$8$\,m telescope with a~$25\%$ throughput and $\lambda_0=800$\,nm). Here, we report results from a parametric study of the~SCC, and propose an easy and robust design to associate it with coronagraphs that use a Lyot stop. We explain the performance in the case of a four-quadrant phase mask coronagraph~\citep[FQPM,][]{Rouan00}. Section\,\ref{sec : sccprin} recalls the principle of the technique and presents the estimators of the pupil complex amplitude~(phase and amplitude errors, step~A) and companion images~(step~B). Section~\ref{sec : assump} provides the assumptions and criterions used for the parametric studies. Section~\ref{sec : DM} is a general study~(no SCC) of one intrinsic limitation for deformable mirrors and presents the best contrast they can provide. The signal-to-noise ratios on both~SCC estimators~(wavefront and companion) are developed in Sect.~\ref{sec : snr}. Section~\ref{sec : reference} estimates the required stability of the reference beam. We report the effects of amplitude aberrations on the~SCC performance in Sect.~\ref{sec : amplitude}. The last two sections are the most important in the paper. The first one is the study of the chromatism on the SCC~performance when a perfect coronagraph is used~(Sect.~\ref{sec : chrom}). In the second one~(Sect.~\ref{sec : sccrealcoro}), we propose a device to associate the SCC and any coronagraph using a Lyot stop. We discuss the case of a~FQPM coronagraph: earths are detected using such a coronagraph with the~SCC just by adding a small hole to the Lyot stop.

    \section{Self-coherent camera principle}
    \label{sec : sccprin}
    The goal of the self-coherent camera is the measurement~(step~A) of phase and amplitude aberrations in the pupil upstream the coronagraph and the speckle field estimation~(step~B) without introducing any non-common path errors. To do so, we use spatial interferences in the science image to encode the stellar speckles that are directly linked to the wavefront aberrations. This section briefly recalls how to use the self-coherent camera to measure wavefront errors and also to reduce the speckle noise in the image.

\begin{figure}[!ht]
      \resizebox{\hsize}{!}{\includegraphics{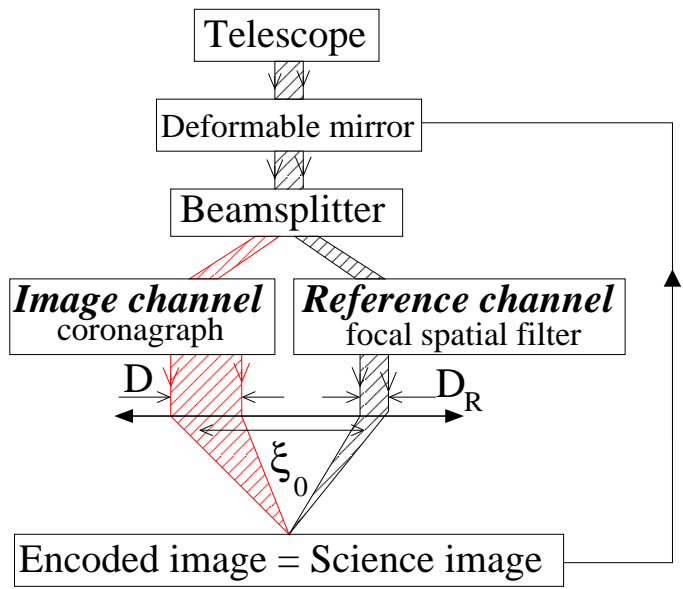}}
      \caption[SCC device schematics]{\it Self-coherent camera principle schematics.}
      \label{fig : schema}
    \end{figure}

\begin{figure}[!ht]
      \resizebox{\hsize}{!}{\includegraphics{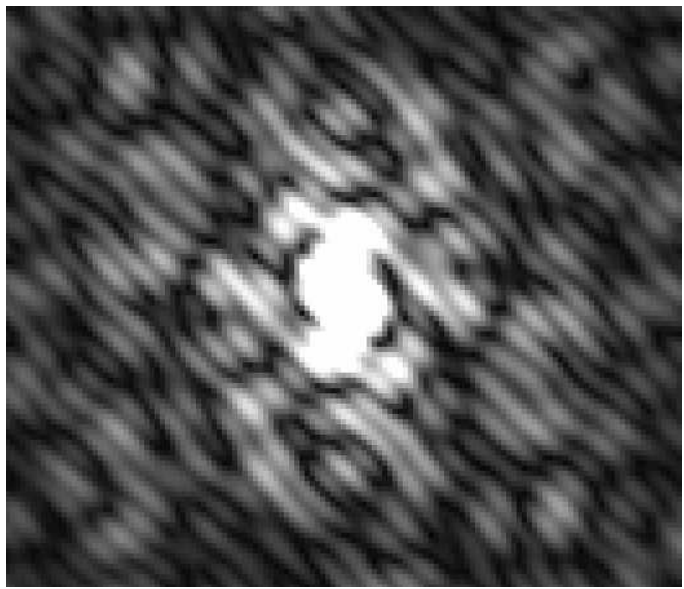}}
      \caption[Science image]{\it Science image of the self-coherent camera. Stellar speckles of the coronagraphic residue are fringed, hence spatially encoded. No companion is present.}
      \label{fig : imageinter}
    \end{figure}

More details can be found in~\citet{Baudoz06,Galicher08}. As in these previous papers, we consider hereafter only space observations. Figure~\ref{fig : schema} presents the instrument schematics.
    A deformable mirror, located in a plane conjugated to the entrance pupil, reflects the beam coming in from the telescope. We then split the beam. In the image channel, the beam goes through a coronagraph. In the reference channel, we suppress all the companion light and extract a beam containing only light from the host star~(cf.~Sect.~\ref{sec : sccrealcoro}). Finally, we recombine the two beams in a Fizeau scheme to obtain spatial fringes in the science image on the detector. Phase and amplitude aberrations give a coronagraphic residue in the last focal plane. And the reference channel induces spatial interferences on these stellar speckles~(cf.~Fig~\ref{fig : imageinter}), whereas it does not have any impact on a possible companion image since companion light is not coherent with star light. The stellar speckles are thus encoded~(modulated), whereas the companion image is not.
We propose the following protocol for detecting faint companions:
\begin{itemize}
\item {\bf Step~A, wavefront estimation and correction}: estimate phase and amplitude errors from the focal plane image and correct for them using a deformable mirror~(correction loop without any differential errors, Sect.~\ref{subsec : stepa})
\item {\bf Step~B, companion detection}: record the science image when the best correction is achieved and post-process that image to overcome the~DM limitation~(Sect.~\ref{subsec : stepb}).
\end{itemize}

    \subsection{Notations}
    Before giving the expression of the wavefront estimator, we summarize the notations of this paper in Table~\ref{tab : notations}. 
    \begin{table}[!ht]
      \renewcommand{\arraystretch}{1.2}
      \begin{center}
        \begin{tabular}{|c|c|}
          \hline
          Wavelength & $\lambda$\\
          \hline
          Central wavelength & $\lambda_0$\\
          \hline
          Bandwidth & $1/R_\lambda = \Delta\lambda/\lambda_0$\\
          \hline
          Star flux in the pupil upstream of the coronagraph & $\psi_0$\\
          \hline
          Pupil diameter  in the image channel & $D$\\
          \hline
          Pupil diameter  in the reference channel & $D_{\mathrm{R}}$\\
          \hline
          $D/D_{\mathrm{R}}$ & $\gamma$\\
          \hline
          Flat pupil of diameter~$D$~($1$ inside, $0$ outside) & $P$\\
          \hline
          Pupil coordinate & $\xi$\\
          \hline
          Phase aberrations upstream the coronagraph & $\phi_\lambda(\xi)$\\
          \hline
          Amplitude aberrations upstream the coronagraph & $a(\xi)$\\
          \hline

   Pupil star amplitude in the image channel & $\psi_{\mathrm{S},\lambda}(\xi)$\\
          \hline
          Pupil star amplitude in the reference channel & $\psi_{\mathrm{R},\lambda}(\xi)$\\
          \hline
          Pupil separation & $\xi_0$\\
          \hline
          Focal plane angular coordinate & $\alpha$\\
          \hline
           Amplitude in the detector plane for $\lambda$ & $A_{i,\lambda}(\alpha)$\\
          \hline
          Intensity in the detector plane for $\lambda$ & $I_{i,\lambda}(\alpha)$\\
          \hline
          Polychromatic intensity in the detector plane & $I_i(\alpha)$\\
          \hline

           Number of actuators of the deformable mirror & $N_{act}$\,x\,$N_{act}$\\
       \hline
            Interferential image in the detector plane  & $I(\alpha)$\\
              \hline
          Estimation of $I_{\mathrm{R}}(\alpha)$ & $\overline{I}_{\mathrm{R}}(\alpha)$\\
            \hline
            Estimation of $A_{\mathrm{R}}(\alpha)$ & $\overline{A}_{\mathrm{R}}(\alpha)$\\
            \hline
        \end{tabular}
        \caption{\it Table of the notations. Index~$i$ represents~$\mathrm{R}$, $\mathrm{S}$, or~$\mathrm{C}$ referring to the reference channel, the stellar residue of the image channel and the companion light of the image channel.}  \label{tab : notations}
      \end{center}
    \end{table}
    Their significance is given in the text when necessary.
   
    \subsection{Step~A: correction loop}
    \label{subsec : stepa}
    \subsubsection{Estimation of the aberrated wavefront}
    \label{subsubsec : estim_phase}
    In~\citet{Galicher08}, we proposed an estimator of phase errors in the pupil upstream of the coronagraph. Here, we give the~SCC estimator of both phase and amplitude aberrations, in other words, the pupil complex amplitude.

    That amplitude is extracted from the science image~$I$~(Fig.~\ref{fig : imageinter}). Using notations from Table~\ref{tab : notations}, we can write the interferential image~$I$ as
    \begin{eqnarray}
      & I(\alpha) = \int_{R_\lambda} \frac{1}{\lambda^2}\Big[\left|A_{\mathrm{S}}(\alpha)\right|^2+\left|A_{\mathrm{R}}(\alpha)\right|^2 +\left|A_{\mathrm{C}}(\alpha)\right|^2 \nonumber \\
        & + 2 Re \left( A_{\mathrm{S}}(\alpha)\,A_{\mathrm{R}}^*(\alpha)\,\exp{\left(\frac{2\,i\,\pi \,\alpha.\xi_0}{\lambda}\right)}\right) \Big]d\lambda,
      \label{eq : intens_focal}
    \end{eqnarray}
where the wavelength $\lambda$ belongs to ${R_\lambda}=[\lambda_0-\Delta\lambda/2,\lambda_0+\Delta\lambda/2]$. The first three terms are the coronagraphic residue~(stellar speckles to be reduced), the reference image, and the companion image~(to be detected). They constitute the unmodulated part of the interferential image. The last term traduces the spatial speckle modulation: interferences with a~$\lambda/\xi_0$ interfringe. The required pupil separation~$\xi_0$~(Fig.~\ref{fig : schema}) not to overlap the peaks in the spatial frequency plane~(inverse Fourier transform of the image~$I$) is at least:
\begin{equation}
\xi_0 = 1.1\,(1.5 + 0.5/\gamma)\,D.
\label{eq : xi_0}
\end{equation}
We use a~$10\%$ margin~(factor~$1.1$).

We follow~\citet{Galicher08} to estimate wavefront errors, and we extract the modulated part~$I_-(\alpha)$ of~$I(\alpha)$ because it contains a linear combination of~$A_{\mathrm{S}}$~(what we look for) and~$A_{\mathrm{R}}$. To do it, we select one of the two lateral peaks of the inverse Fourier transform of~$I$ and apply a Fourier transform to it:
\begin{equation}
  I_-(\alpha) = \int_{R_\lambda} \frac{1}{\lambda^2}\,A_{\mathrm{S}}(\alpha)\,A_{\mathrm{R}}^*(\alpha)\,\exp{\left(\frac{2\,i\,\pi\,\alpha.\xi_0}{\lambda}\right)}\,d\lambda.
\label{eq : i_-def}
\end{equation}
We then assume a small effective bandwidth~(cf.~Sect.~\ref{sec : chrom}) and estimate the pupil complex amplitude downstream the coronagraph:
\begin{equation}
  \psi_{\mathrm{S}}(u) \simeq \mathcal{F}^{-1}\left[\displaystyle \frac{I_-(\alpha)}{A_{\mathrm{R}}^*(\alpha)\,F(\alpha)}\right](u),
  \label{eq : psiS_0}
\end{equation}
where~$\mathcal{F}^{-1}$ is the inverse Fourier transform, $u$ the cooordinate in the correlation plane of~$I$, and $F$ is applied to partially compensate for chromatic effects~(see Sect.~\ref{subsec : factor}). Assuming small phase~($\phi$) and amplitude~($a$) aberrations and a perfect coronagraph, we find the complex amplitude~$\psi_{\mathrm{S}}$ downstream of the coronagraph~\citep[see Eqs.~$6$ to~$8$ in][for details]{Galicher08}:
\begin{equation}
\psi_{\mathrm{S}}(u) = \psi_0\,[a(u)+i\,\phi(u)]\,P(u),
\end{equation}
where~$psi_0$ is the star flux upstream of the coronagraph. From the two last equations, we deduce
\begin{equation}
  \phi(u) -i\,a(u) \simeq -i\,\mathcal{F}^{-1}\left[\displaystyle \frac{I_-(\alpha)}{\psi_0\,A_{\mathrm{R}}^*(\alpha)\,F(\alpha)}\right](u).
\label{eq : phase_est}
\end{equation}
We can thus estimate wavefront errors upstream of the coronagraph from~$I_-$, extracted from the science image. The other terms can be estimated as follows. The chromatic factor is computed, and its expression is given in Sect.~\ref{subsec : factor}. The constant~$\psi_0$ is uniform~(Table~\ref{tab : assumptions}). We can precisely estimate it from flux calibration of the space telescope. Finally, the reference amplitude in the focal plane~$A_{\mathrm{R}}^*$ has to be recorded before the beginning of the loop and has to be stable. We point out that such a stabilization can be achieved~(Sect.~\ref{sec : reference}).\\ Once wavefront aberrations are estimated, we correct for them with a deformable mirror as explained in the next section.

\subsubsection{Correction of the aberrated wavefront}
\label{subsubsec : DMcorrection}
In the whole paper, we assume that we use only one deformable mirror with~$N_{\mathrm{act}}$x$N_{\mathrm{act}}$ actuators to correct for the wavefront errors. If only phase aberrations exist, it is possible to clean all the~$\left([-N_{\mathrm{act}}/2,N_{\mathrm{act}}/2]\,\left(\lambda_0/D\right)\right)^2$ centered area in the science image. If both phase and amplitude aberrations are present, we can only clean half of that region, and  we adopt the method proposed by~\citet{Borde06} to make the phase error screen hermitian. As the estimation is not perfect~(small aberration linearization, noises, reference division) and the speckles are not static but quasi-static, we need to iterate a few times.

To achieve high-contrast imaging, we need to measure wavefront aberrations with high accuracy and to drive the deformable mirror faster than the aberration evolution time. As the required time to work out the~SCC estimator is relatively short~(three~fast Fourier transform), the correction speed is determined by the integration time needed to reach a reasonable signal-to-noise ratio for the wavefront measure. There is then a compromise between the achievable image contrast and the pointed stellar flux. That compromise is roughly the same regardless of the speckle calibration technique. Recent industrial studies for a space-based~$1.5$m telescope estimate that stabilities can be expected as low as~$1\,$pm per hour~\citep{Guyon09}.

\subsection{Step~B: companion estimation}
\label{subsec : stepb}
Once the best correction is achieved~(last iteration of step~A), we apply the post-proccessing algorithm described in~\citet{Baudoz06} and~\citet{Galicher07} to suppress the largest part of the speckle residue. We estimate the companion image~$I_{\mathrm{C\,est}}$ using
\begin{equation}
I_{\mathrm{C\,est}}(\alpha)=I_{\mathrm{cent}}(\alpha)-I_{\mathrm{R}}(\alpha)-\frac{|I_{-}(\alpha)|^2}{I_{\mathrm{R}}(\alpha)},
\label{eq : est_comp}
\end{equation}
where~$I_{-}$ is defined by~Eq.~\ref{eq : i_-def} and~$I_{\mathrm{cent}}$ is the unmodulated part of the science image~$I$~(Eq.~\ref{eq : intens_focal}), which can be written as
\begin{equation}
  I_{\mathrm{cent}}(\alpha) = \int_{R_\lambda} \frac{1}{\lambda^2}\Big[I_{\mathrm{S}}(\alpha)+I_{\mathrm{R}}(\alpha) +I_{\mathrm{C}}(\alpha)\Big]\,d\lambda.
\label{eq : icent_def}
\end{equation}
To extract it, we select the central peak of the inverse Fourier transform of the science image~(Eq.~\ref{eq : intens_focal}) and apply a Fourier transform to that selection. Finally, $I_{\mathrm{R}}=|A_{\mathrm{R}}|^2$ is the reference image. It has to be estimated as explained in the introduction of Sect.~\ref{sec : reference}.

\section{Assumptions and criteria}
\label{sec : assump}
This section introduces the assumptions of our numerical studies and the criteria for optimizing all the parameters of the instrument.
\subsection{Assumptions}
\label{subsec : assumptions}
In the whole paper, we assume spatial observations~(no dynamic aberrations) and achromatic coronagraphs without defects~(perfect coronagraph or FQPM). The power spectral density~(PSD) of static phase errors in the instrument upstream of the coronagraph varies as~$f^{-3}$, where~$f$ is the spatial frequency, which corresponds to typical VLT mirror aberrations~\citep{Borde06}. Amplitude aberration~PSD is flat or evolves as~$f^{-3}$.
\begin{table*}[!ht]
  \renewcommand{\arraystretch}{1.2}
  \begin{center}
    \begin{tabular}{|c|c|c|c|c|c|c|c|c|c|c|}
      \hline
      \backslashbox{parameter}{section} & \ref{sec : DM} & \ref{subsec : gamma}&\ref{subsec : opd}& \ref{sec : amplitude}&\ref{subsec : factor} & \ref{subsec : chromlimit}& \ref{para : ref_flux_fqpm}& \ref{para : sccfqpm_chrom} & \ref{para : sccfqpm_detect}\\
      \hline
      SCC & no & yes& yes& yes& yes& yes& yes& yes& yes\\
      \hline
      $\gamma$ & \slashbox{}{} & $20$ & $1000$& $1000$& $1000$& $1000$& $10$ and~$20$ & $20$&$25$\\
      \hline
      $R_\lambda$ & $\infty$ &$\infty$&$\infty$&$\infty$ & $40$ & $16$ to $\infty$ & $\infty$ & $16$ to $\infty$ & $150$\\
      \hline
      $N_{\mathrm{act}}$ & $16$, $32$, $64$  &$32$&$32$&$64$&$64$ & $16$, $32$, $64$ & $32$ & $32$ &$64$\\
       \hline
      coronagraph & perfect& perfect& perfect& perfect& perfect& perfect &FQPM&FQPM&FQPM\\
      \hline
      $\sigma(\phi)$\,rms~(nm) &  $20$& $20$& $20$& $20$& $20$& $20$ & $20$ and~$40$& $20$& $20$\\
      \hline
       &  & & & $0$ to $1$  & & & & & $1$\\
 \multirow{-2}{*}{$\sigma(a)$\,rms~($\%$)} &\multirow{-2}{*}{$0$}&\multirow{-2}{*}{$0$}&\multirow{-2}{*}{$0$}& ($f^{-3}$ PSD) &\multirow{-2}{*}{$0$}&\multirow{-2}{*}{$0$}&\multirow{-2}{*}{$0$}&\multirow{-2}{*}{$0$}&(flat PSD)\\
      \hline
      Photon noise & no &no&no&no&no&no &no & no & yes\\
      \hline
    \end{tabular}
    \caption{\it Simulation assumptions for each section. Notations are defined in Table~\ref{tab : notations}. Details as photon number are given in the text.}
\label{tab : assumptions}
  \end{center}
\end{table*}
The reference complex amplitude~$A_\mathrm{R}$ in the detector plane is the Fourier transform of
\begin{itemize}
\item the pupil upstream of the coronagraph~(same wavefront errors) densified to obtain a pupil diameter~$D_{\mathrm{R}}$ instead of~$D$, if a perfect coronagraph is used. The flux is set to verify the condition of Sect.~\ref{subsec : irmin} in the corrected area.
\item a pupil of diameter~$D_{\mathrm{R}}$ extracted from the Lyot stop plane~(phase and amplitude errors depend on the coronagraph) if a FQPM coronagraph is used~(see Sect.~\ref{sec : sccrealcoro}).
\end{itemize}

We consider the deformable mirror as a continous face sheet supported by~$N_{act}$\,x\,$N_{act}$ actuators arranged in a square pattern of constant spacing~\citep{Borde06} and located in a plane conjugate to the entrance pupil. The nth-actuator influence function is~$\exp{(\ln{(0.15)}\,(N_{act}\,(\xi-\xi_{\mathrm{n}})/D)^2)}$, where $\xi_{\mathrm{n}}$ is the center of the nth-actuator~\citep{huang08}. We use the method of {\it energy minimization in the pupil plane}~\citep[Eq.~$A3$ of ][]{Borde06} to project the estimated phase on the deformable mirror and shrink the corrected area by a~$1.05$ factor to optimize the correction.

To simulate the science image~$I$~(cf.~Eq.~\ref{eq : intens_focal}) in polychromatic light, we sum $5$ monochromatic images corresponding to~wavelengths uniformly distributed in the considered bandpass~$\Delta\lambda=\lambda_0/R_\lambda$. Shannon sampling is imposed for fringes~$(\lambda_0/\xi_0)_{\mathrm{pix}}=4$, and~$\xi_0$ follows~Eq.~\ref{eq : xi_0}. When photon noise is simulated, we consider a~G$2$ star at~$10$\,pc observed by a space telescope with~$40\%$ of throughput and a bandwidth of~$20\%$ at~$\lambda_0=800$\,nm. We indicate the exposure time and the telescope diameter case by case. Table~\ref{tab : assumptions} regroups the other assumptions, which change from one section to another. Only one parameter generally takes realistic values so that its impact can be determined independently to the other parameters, set to ideal values. At the end of Sect.~\ref{sec : sccrealcoro}, all parameters are set to realistic values to determine the~SCC-FQPM performance.

\subsection{Criteria}
\label{subsec : criterion}
To optimize the different parameters of the self-coherent camera, we define two criteria.
\subsubsection{Averaged contrast}
The first criterion, called~$C_1$, gives the averaged contrast achieved in the corrected area~$\mathcal{H}$ of the SCC science image. This area varies as a function of three parameters : deformable mirror size, polychromatism and amplitude aberrations. Considering an~$N_{act}$x$N_{act}$ deformable mirror:
\begin{itemize}
\item in monochromatic light and without amplitude aberrations, $\mathcal{H}$ is the centered~$(0.9/1.05\,N_{act})$x$(0.9/1.05\,N_{act})$\,$(\lambda_0/D)^2$ region because we reduce the corrected area by shrinking it by a factor~$1.05$~(Sect.~\ref{subsec : assumptions}) and do not account for borders~(factor~$0.9$) of that area where the detection is prevented by the diffracted light of the uncorrected speckles.
\item in polychromatic light and without amplitude aberrations, following results of Sect.~\ref{subsec : fieldlimit}, we reduce~$\mathcal{H}$ to its intersection with the line of width~$\alpha_{\mathrm{B}}$ and of same direction as fringes.
\end{itemize}
If amplitude aberrations exist, we consider the region where the correction is effective~(Sect.~\ref{sec : amplitude}), which corresponds to half the~$\mathcal{H}$ area defined herebefore. The~$C_1$ can be written as
\begin{equation}
 C_1 = \frac{\int_{\mathcal{H}}I(\alpha)\mathrm{d}\alpha}{I_0\int_{\mathcal{H}}\mathrm{d}\alpha},
 \label{eq : c_1}
\end{equation}
where $I_0$ is the maximum intensity of the star image without coronagraph or SCC, and~$I$ is given by~Eq.~\ref{eq : intens_focal}. The smaller $C_1$, the better the correction and the fainter companions can be detected. Since $C_1$ is an average over the corrected area, it does not give any information on preferential positions in the image for the detection.

As explained in~Sect.~\ref{subsubsec : DMcorrection}, several iterations are required to reach very high-contrast imaging. That is why, when~$C_1$ is considered, we study its convergence speed as a function of the number of correction steps.

\subsubsection{$5\,\sigma$ detection}
The second quantity we use to compare the different configurations is the~$5\,\sigma$ detection~$d_{5\,\sigma}$, which we define as
\begin{equation}
  d_{5\,\sigma}(\rho) = \frac{5\,\sigma(\rho)}{I_0},
\end{equation}
where $\sigma(\rho)$ is the azimuthal standard deviation of the considered image at radial separation~$\rho$. As for~$C_1$, the smaller $d_{5\,\sigma}$, the better the correction and the fainter companions can be detected. The~$d_{5\,\sigma}$ metric is a function of the angular separation in the considered image. It is then possible to determine preferential angular separations for the detection. However, $d_{5\,\sigma}(\rho)$ does not give any information on favored directions. If $d_{5\,\sigma}(\rho)$ is used, we plot it against the angular separation.

\section{Intrinsic deformable mirror limitation}
\label{sec : DM}
\begin{figure}[!ht]
  \resizebox{\hsize}{!}{\includegraphics{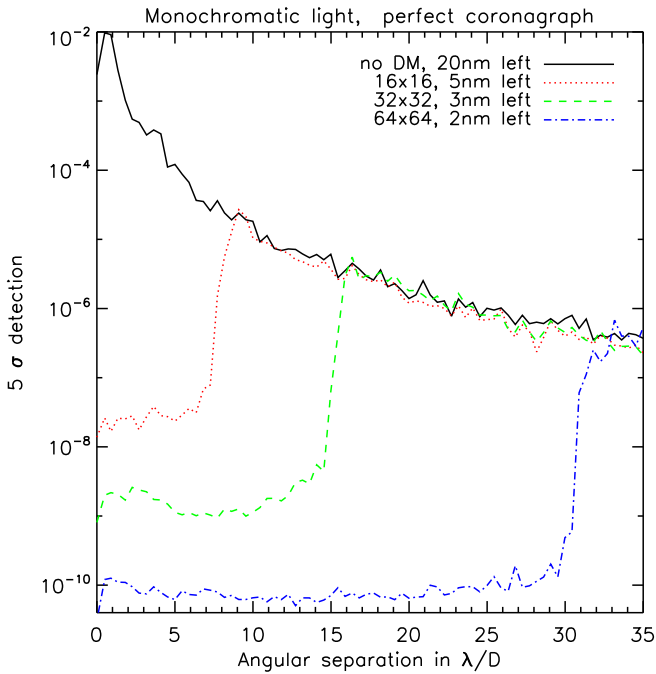}}
  \caption[DM correction limit]{\it $5\,\sigma$ detection versus angular separation in the image provided by a perfect coronagraph associated to a single telescope in monochromatic light without DM~(solid black line) or with DM: $16$x$16$, $32$x$32$ and $64$x$64$ from top to bottom. The initial phase error rms amplitude is~$20$\,nm with a power spectral density in~$f^{-3}$. Residual phase error levels are given for each~DM in~nm.}
  \label{fig : dmlimitsize}
\end{figure}

To derive SCC performance, we needed to know the intrinsic limitation of deformable mirrors that we use. In that section, we set out that intrinsic limitation~(no~SCC) if no apodization is used. We considered a single telescope associated with a perfect coronagraph and a deformable mirror. All the assumptions are recalled in Sect.~\ref{subsec : assumptions}. We simulated the coronagraphic image, without SCC, with a full DM correction~(perfect estimation of the phase errors). We ploted~$d_{5\sigma}$ against the angular separation~$\rho$ in Fig.~\ref{fig : dmlimitsize} for three different~DM sizes and the same phase screen. The residue after the perfect coronagraph~(no~DM) is overplotted.

As expected, the cutoff frequency~$\rho_c$ equals~$(N_{act}/2)\,(\lambda_0/D)$, and higher frequencies are not corrected: no change for~$d_{5\sigma}(\rho)$ with or without correction further than~$\rho_c$.
For $\rho<\rho_c$, the detection limit depends on the diffracted energy of uncorrected speckles located outside the corrected area~($\rho>\rho_c$). The larger~$N_{act}$, the larger~$\rho_{\mathrm{C}}$ and the lower the level of uncorrected speckles. That is why, for a fixed~$\rho<\rho_{\mathrm{C}}$, the correction is better with a $64$x$64$~DM than with a $32$x$32$.

To improve the detection limit we can increase the number of actuators~(degrees of freedom of the correction). However, nowadays, manufacturing issues limit that number to~$64$x$64$ for MicroElectroMechanical Systems~(MEMS). For the E-ELT planet finder, EPICS, a~$200$x$200$ device is under study but has not been developed yet. Another way to improve the detection limit is to minimize wavefront errors but it would be very optimistic to assume less than~$20$\,nm~rms for telescopes larger than a few meters. Another solution to rule out the detection limit is apodize the pupil so that the uncorrected speckles would spread less light into the corrected area. \citet{giveon06} assume a Kaiser function to apodize the pupil and achieve a~$10^{-10}$ contrast~($1\,\sigma$ detection) at~$4\,\lambda_0/D$ in the focal plane. But such an apodization profile is are not easy to produce. With an amplitude mask~\citep{vanderbei03,kasdin05}, its throughput would be about~$15\%$, which significantly increases the exposure time for detecting faint companions. Another apodization technique, called phase-induced amplitude apodization~\citep{Guyon05}, could be used because its throughput is about~$100\%$. But manufacturing problems limit its performance, even if solutions are under study~\citep{Pluzhnik06}.

Finally, if no apodization is used, the best~DM correction shown in~Fig.~\ref{fig : dmlimitsize}
\begin{itemize}
\item cannot be overcome by any device used to estimate wavefront errors because it is an intrinsic limitation depending on the diffraction of uncorrected speckles;
\item comes from a speckle residue so that the sole way to improve it is a post-correction of the image by differential imaging techniques. One example is the self-coherent camera~(step~B, Sect.~\ref{subsec : stepb}).
\end{itemize}

We provide below results of a parametric study of the self-coherent camera used as a focal plane wavefront sensor.

\section{Signal-to-noise ratio}
\label{sec : snr}
We study the impact of the reference flux on the signal-to-noise ratio of wavefront error estimation~(step~A, Sect.~\ref{subsec : irmin}) and companion detection~(step~B, Sect.~\ref{subsec : icmin}).

\subsection{Step A: wavefront error measurement}
\label{subsec : irmin}
The purpose of step~A is to measure wavefront errors directly from the noisy science image~$I$. The interesting quantity is then~$A_{\mathrm{S}}$ and, here, we work out the signal-to-noise ratio of that measure considering the photon noise.

To estimate~$A_{\mathrm{S}}$, we use~Eq.~\ref{eq : i_-def} and obtain in monochromatic light
\begin{equation}
A_{\mathrm{S,n}} = \frac{I_{\mathrm{-,n}}}{A^*_{\mathrm{R}}},
\label{eq : as_est}
\end{equation}
where the~$n$ index refers to the noisy quantities and where~$I_-$ is the Fourier transform of one of the two lateral correlation peaks of the science image~$I$. Assuming~$A^*_{\mathrm{R}}$ is almost uniform in the image~(see Sect.~\ref{sec : reference}), Eq.\ref{eq : as_est} implies
\begin{equation}
Var[A_{\mathrm{S,n}}] = \frac{Var[I_{\mathrm{-,n}}]}{I_{\mathrm{R}}},
\label{eq : as_var}
\end{equation}
where we have replaced~$|A_{\mathrm{R}}|^2$ by the intensity of the reference channel on the detector~$I_{\mathrm{R}}$. Considering photon noise, variance of~$I_{\mathrm{-,n}}$ can be approximated by~(see Eq.~\ref{eq : var_epsilon_-_B} of Appendix~A):
\begin{equation}
Var[I_{\mathrm{-,n}}] \simeq \frac{I_{\mathrm{S}} +I_{\mathrm{R}}}{(\lambda_0/D)^2_{\mathrm{pix}}},
\label{eq : i-_var}
\end{equation}
where~$(\lambda_0/D)_{\mathrm{pix}}$ is the number of pixels per resolution element~$\lambda_0/D$ in the science image. From~Eqs.~\ref{eq : as_var} and~\ref{eq : i-_var}, if the~$A_{\mathrm{S,n}}$ measure is done with a signal-to-noise ratio higher than~$\eta$, we can write
\begin{equation}
\frac{|I_{\mathrm{-,n}}|}{|A^*_{\mathrm{R}}|}\,\sqrt{\frac{I_{\mathrm{R}}}{I_{\mathrm{S}} +I_{\mathrm{R}}}}\,\left(\frac{\lambda_0}{D}\right)_{\mathrm{pix}}>\eta.
\end{equation}
Supplanting~$|I_-|$ by~$|A_{\mathrm{S}}\,A^*_{\mathrm{R}}|=\sqrt{I_{\mathrm{S}}\,I_{\mathrm{R}}}$~(Eq.~\ref{eq : as_est}) and~$|A^*_{\mathrm{R}}|$ by~$\sqrt{I_{\mathrm{R}}}$, we find:
\begin{equation}
\sqrt{\frac{I_{\mathrm{S}}\,I_{\mathrm{R}}}{I_{\mathrm{S}} +I_{\mathrm{R}}}}\,\left(\frac{\lambda_0}{D}\right)_{\mathrm{pix}}>\eta.
\end{equation}
We want to estimate for wavefront errors~(directly linked to~$I_{\mathrm{S}}$) and we know the reference intensity~$I_{\mathrm{R}}$. In the case of an association with a~FQPM~(Sect.~\ref{para : ref_flux_fqpm}), $I_{\mathrm{S}}\gg I_{\mathrm{R}}$ before beginning the correction loop. Under that condition,~$\eta_{\mathrm{max}}\sim(\lambda_0/D)_{\mathrm{pix}}\,\sqrt{I_{\mathrm{R}}}$. Using the Shannon fringe sampling~($(\lambda_0/\xi_0)_{\mathrm{pix}}\simeq 3$) and the minimum separation~$\xi_0$ between pupils of image and reference channels~(Eq.~\ref{eq : xi_0}), the signal-to-noise ratio at the beginning of the loop roughly equals~$5\,\sqrt{I_{\mathrm{R}}}$. After a few correction steps, we hope to achieve~$I_{\mathrm{S}}\sim I_{\mathrm{R}}$. In that case,~$\eta_{\mathrm{max}}\sim(\lambda_0/D)_{\mathrm{pix}}\,\sqrt{I_{\mathrm{R}}/2}\sim 3.5\,\sqrt{I_{\mathrm{R}}}$. Finally, wavefront estimation~(step~A) can be done with a good accuracy as soon as~$I_{\mathrm{R}}$ is greater than a few photons per pixel. The sole problem would be a zero value of the reference flux because the speckles would not be encoded and~$\eta$ would be zero~(see Sect.~\ref{sec : reference}).

\subsection{Step~B: companion detection}
\label{subsec : icmin}
Step~B corresponds to the companion estimation~$I_{\mathrm{C\,est}}$ from the science image~$I$ of the last iteration of the correction~(end of step~A). It would be quite exhausting to study the exact propagation of noises through all steps of the algorithm that we use to compute the estimator of~Eq.~\ref{eq : est_comp}. We may present such a study in a future paper but it is not the purpose of the present one. Here, we express the dependence of the signal-to-noise ratio of the estimator on fluxes in the image channel~($I_{\mathrm{S}}$, speckle noise) and in the reference channel~($I_{\mathrm{R}}$).

We first assume uncorrelated noises for the unmodulated~($I_{\mathrm{cent,n}}$) and modulated~($|I_{\mathrm{-,n}}|^2$) parts of the science image~$I$~(not rigorously exact). Considering the recorded reference intensity~$\overline{I}_{\mathrm{R}}$ is almost uniform on the detector~(see Sect.~\ref{sec : reference}), we can write the variance of the estimator~$I_{\mathrm{C\,est,n}}$~(Eq.~\ref{eq : est_comp}):
\begin{equation}
Var[I_{\mathrm{C\,est,n}}] \simeq Var[I_{\mathrm{cent,n}}] + \frac{Var[|I_{\mathrm{-,n}}|^2]}{\overline{I}^2_{\mathrm{R}}}.
\label{eq : noise_icest0}
\end{equation}
We restrict the covariance of~$I_{\mathrm{-,n}}$ to its variance~(see~Eq.~\ref{eq : covar_epsilon_-bis} of Appendix~A) and we suppose the distribution of the noise on~$I_{\mathrm{-,n}}$ still follows a Poissonian law. We can then write the fourth momentum of~$I_{\mathrm{-,n}}$:
\begin{equation}
Var[|I_{\mathrm{-,n}}|^2] \simeq Var[|I_{\mathrm{-,n}}|] + 3\,Var[|I_{\mathrm{-,n}}|]^2.
\label{eq : fourth_momentum_i-}
\end{equation}
Finally, using~Eqs.~\ref{eq : var_epsilon_-_B} and~\ref{eq : var_icent_B} of Appendix~A~(variances of~$I_{\mathrm{-,n}}$ and~$I_{\mathrm{cent,n}}$) and~Eqs.~\ref{eq : noise_icest0} and~\ref{eq : fourth_momentum_i-}, we obtain an approximation of the noise on~$I_{\mathrm{C\,est,n}}$:
\begin{equation}
Var[I_{\mathrm{C\,est,n}}] \simeq I(\alpha)\,\left(\frac{D}{\lambda_0}\right)^2_{\mathrm{pix}}\,\left(4 + \frac{1}{\overline{I}^2_{\mathrm{R}}} + \frac{3\,I(\alpha)}{\overline{I}^2_{\mathrm{R}}} \left(\frac{D}{\lambda_0}\right)^2_{\mathrm{pix}}\right),
\label{eq : noise_icest}
\end{equation}
where~$I$ roughly equals to~$I_{\mathrm{S}}+I_{\mathrm{R}}+I_{\mathrm{C}}$. We insist that this expression is not exact, but we did numerically check that it roughly approximates the real noise. We can then derive the ideal flux regime~($I_{\mathrm{S}}$ and~$I_{\mathrm{R}}$) to optimize the speckle substraction when applying estimator of~Eq.~\ref{eq : est_comp}. To do it, we assume the companion intensity is low compared to the coronagraphic residue~($I_{\mathrm{C}}\ll I_{\mathrm{S}}$). If this is not the case, the companion is detected and the~SCC not useful.\\

{\bf Weak fluxes\\}
If~$I_{\mathrm{C}}\ll I_{\mathrm{S}}\simeq I_{\mathrm{R}}\ll 1$\,photon per pixel, the variance of the noise on~$I_{\mathrm{C\,est,n}}$ reduces to
\begin{equation}
Var[I_{\mathrm{C\,est,n}}] \simeq \frac{2}{\overline{I}_{\mathrm{R}}}\,\left(\frac{D}{\lambda_0}\right)^2_{\mathrm{pix}},
\end{equation}
and the signal-to-noise ratio~$\beta$ of the companion detection is
\begin{equation}
\beta \simeq I_{\mathrm{C}}\,\sqrt{\frac{\overline{I}_{\mathrm{R}}}{2}}\,\left(\frac{\lambda_0}{D}\right)_{\mathrm{pix}}\ll 1,
\end{equation}
which is not interesting.\\

{\bf Dominating coronagraphic residue\\}
If~$I_{\mathrm{C}}\ll I_{\mathrm{S}}$, $1\lesssim \ll I_{\mathrm{S}}$ and~$I_{\mathrm{R}}\ll I_{\mathrm{S}}$, Eq.~\ref{eq : noise_icest} becomes
\begin{equation}
Var[I_{\mathrm{C\,est,n}}] \simeq \frac{3\,I^2_{\mathrm{S}}}{\overline{I}^2_{\mathrm{R}}}\,\left(\frac{D}{\lambda_0}\right)^4_{\mathrm{pix}}\gtrsim I^2_{\mathrm{S}}.
\end{equation}
If~$I_{\mathrm{R}}\lesssim 1$, noise on the companion estimator is greater than the initial speckle noise and the~SCC is useless. If~$I_{\mathrm{R}}\gtrsim1$, speckle noise is lightly reduced and the~SCC acts positively.\\

{\bf Dominating reference image\\}
If~$I_{\mathrm{C}}\ll I_{\mathrm{S}}\ll I_{\mathrm{R}}$ and~$1\ll I_{\mathrm{R}}$, we simplify~Eq.~\ref{eq : noise_icest} as
\begin{equation}
Var[I_{\mathrm{C\,est,n}}] \simeq 4\,I_{\mathrm{R}}\,\left(\frac{D}{\lambda_0}\right)^2_{\mathrm{pix}}\varpropto I_{\mathrm{R}}.
\end{equation}
The reference-image photon noise dominates the estimator noise. As it is greater than the speckle residue that we want to reduce~($I_{\mathrm{S}}\ll I_{\mathrm{R}}$), this case is not interesting again.\\

{\bf Strong fluxes for both channels\\}
If~$I_{\mathrm{C}}\ll I_{\mathrm{S}}$ and~$I_{\mathrm{R}}\simeq I_{\mathrm{S}}\gg1$\,photon per pixel, the variance of the noise on the companion estimator~$I_{\mathrm{C\,est,n}}$~(Eq.~\ref{eq : noise_icest}) reduces to
\begin{equation}
Var[I_{\mathrm{C\,est,n}}] \simeq 8\,I_{\mathrm{S}}\,\left(\frac{D}{\lambda_0}\right)^2_{\mathrm{pix}}\varpropto I_{\mathrm{S}},
\end{equation}
which means that it is proportional to the photon noise of the coronagraphic residue~(of which the variance is~$I_{\mathrm{S}}+I_{\mathrm{R}}\simeq2\,I_{\mathrm{S}}$): the speckle noise is suppressed. The signal-to-noise ratio~$\beta$ of the companion detection is then
\begin{equation}
\beta \simeq \frac{I_{\mathrm{C}}}{8\,I_{\mathrm{S}}}\,\left(\frac{\lambda_0}{D}\right)_{\mathrm{pix}}\simeq\frac{3\,I_{\mathrm{C}}}{I_{\mathrm{S}}}.
\end{equation}
We have assumed~$(\lambda_0/D)_{\mathrm{pix}}\simeq5$ to verify the Shannon sampling of fringes~($(\lambda_0/\xi_0)_{\mathrm{pix}}\simeq3$) and the minimum separation between image and reference pupils~(Eq.~\ref{eq : xi_0}). The factor~$3$ reveals that the noise is smoothed and averaged when we select the correlation peaks and work out~$I_{\mathrm{-,n}}$ and~$I_{\mathrm{cent,n}}$ by fast Fourier transform.\\

\vspace{.1cm}
Finally, the best way to estimate the companion image from~Eq.~\ref{eq : est_comp} with a good signal-to-noise ratio is to set strong fluxes in both image and reference channels~($I_{\mathrm{S}}$ and $I_{\mathrm{R}}$ larger than~$1$\,photon per pixel on the detector).

For both wavefront estimation~(step~A) and companion estimation~(step~B), the range of accepted reference fluxes~$I_{\mathrm{R}}$ is broad and we propose to work with~$I_{\mathrm{R}}$ around a few tens of photons per pixel in the science image.

\section{Reference beam}
\label{sec : reference}
Like all differential imaging techniques, SCC needs the recording of a reference whose the complex amplitude~$A_{\mathrm{R}}$ is used to estimate wavefront errors~(Eq.~\ref{eq : phase_est}). A first problem~(zero division) appears at the positions where~$A_{\mathrm{R}}$ is zero because the condition of Sect.~\ref{subsec : irmin} is not verified and the corresponding speckles are not encoded well. The second problem is that it is impossible to simultaneously measure~$A_{\mathrm{R}}$ and the science image. We then propose to record the reference complex amplitude before beginning the loop and we call it~$\overline{A}_{\mathrm{R}}$. But in that case we need that recording to be stable in time. In Sect.~\ref{subsec : gamma}, we present a way to answer both problems at the same time. Then, we study the impact of errors on~$\overline{A}_{\mathrm{R}}$: spatial drift~\ref{subsec : gamma} and optical path difference variation~(Sect.~\ref{subsec : opd}) between the recording of~$\overline{A}_{\mathrm{R}}$ and the beginning of the loop.

It is important to distinguish the reference amplitude~$A_{\mathrm{R}}$ used to simulate the~SCC science fringed image~(see Sect.~\ref{subsec : assumptions}) and the reference amplitude~$\overline{A}_\mathrm{R}$ used in the estimator of~Eq.~\ref{eq : phase_est}. The second one is an estimation of the {\it a priori} unknown first one. In this paper, we use the diffracted complex amplitude by a $D_{\mathrm{R}}$-diameter pupil free from any aberrations for~$\overline{A}_\mathrm{R}$. Flux and position of~$\overline{A}_\mathrm{R}$ are set using a recorded image of the reference channel before the beginning of the correction loop.

\subsection{Reference pupil diameter and spatial drifts of the image}
\label{subsec : gamma}
The solution proposed to stabilize the reference channel and avoid the zero divisions is to use a small diameter~$D_{\mathrm{R}}$ for the reference pupil. In that way,~$A_{\mathrm{R}}$, roughly equal to the central part of the Airy complex amplitude, has very low sensitivity to wavefront variations in the reference channel, and its first dark ring is pushed away from the center of the image. At this point, we would like to set the smallest~$D_{\mathrm{R}}$ that we could. However, in a real setup, the light entering the reference channel is extracted from the coronagraph rejected light, and its flux has a finite value. A minimum value for~$D_{\mathrm{R}}$ is then required to verify the condition of Sect.~\ref{subsec : irmin} and a trade-off has to be derived for each setup. In the case of an association with a~FQPM coronagraph~\citep{Rouan00}, we find that a reasonable value of~$\gamma=D/D_{\mathrm{R}}$ is between~$\sim10$ to~$\sim30$ in function of wavefront error level~(Sect.~\ref{subsec : sccfqpm}), and we often use~$\gamma=20$ in the paper.

We now quantify the impact of errors on the recorded reference amplitude~$\overline{A_{\mathrm{R}}}$. We first consider tip-tilt variations in the reference channel that induce spatial drifts of the reference image on the detector between~$\overline{A_{\mathrm{R}}}$ and~$A_{\mathrm{R}}$. With~$\gamma=20$ and under the assumptions mentioned in Sect.~\ref{subsec : assumptions}, we find that for a~$\sim5\,\lambda_0/D$~(respectively~$6$) drift of the reference image, the correction is effective and roughly converges to the~DM limitation~(twice the~DM limitation). The specification is then not critical, and in Sect.~\ref{sec : sccrealcoro}, we propose a compact, robust, and very simple device where the reference is stable enough in time.

\subsection{Optical path difference}
\label{subsec : opd}
\begin{figure}[!ht]
    \resizebox{\hsize}{!}{\includegraphics{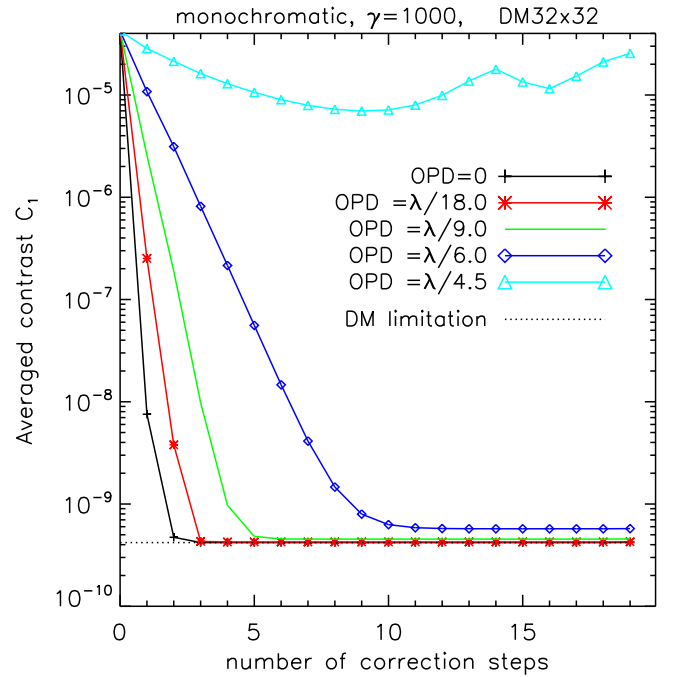}}
    \caption[OPD and perfect coronagraph]{\it Averaged contrast~$C_1$ in the corrected area against correction steps for different optical path differences~(OPD) between reference and image channels. We assume a~$32$x$32$ deformable mirror.}
    \label{fig : opd_parf}
  \end{figure}

Another potential error on the estimated reference amplitude~$\overline{A_{\mathrm{R}}}$ is the variation of the optical path difference~(OPD) between the image and the reference channels. If the~OPD varies, the center of the fringe pattern is shifted on the detector and, if we do not account for it, the wavefront error estimation is degraded. To quantify that limitation, we consider that the recorded reference~$\overline{A_{\mathrm{R}}}$ corresponds to a zero~OPD and we simulate correction loops for several non-zero~OPD under assumptions of Sect.~\ref{subsec : assumptions}. Figure~\ref{fig : opd_parf} gives the convergence of the averaged contrast~$C_1$ in the corrected area.
If the~OPD is zero, the correction reaches the DM limitation~(Sect.~\ref{sec : DM}) in~$3$ steps. When inscreasing the~OPD, the correction converges more and more slowly. If the~OPD is around the critical value~$\lambda_0/4$, the correction does not converge anymore. In that case, the estimated reference amplitude~$\overline{A_{\mathrm{R}}}$~(used in the wavefront estimator) equals to the real reference amplitude~$A_{\mathrm{R}}$ multiplied by~$i$~(imaginary unit). All phase errors are estimated as amplitude errors and vice versa. The correction is not effective. To have the correction converging close to the DM limitation, we need to know the~OPD between reference and image channels with an accuracy of~$\sim\lambda_0/6$. And we have to stabilize it with the same precision~($\sim\lambda_0/6$) during the correction loop. It is possible to determine the~OPD with accuracy just before closing the loop by recording a science image with a large bandwidth. The white fringe corresponds to the zero optical path difference, and its spatial shift from the optical axis is directly linked to the~OPD. To ensure the stability, we propose a robust compact device in the case of the association of the~SCC with a coronagraph using a Lyot stop as the~FQPM coronagraph~(Sect.~\ref{para : fqpm_opd}).

\section{Amplitude aberrations}
\label{sec : amplitude}
\begin{figure}[!ht]
  \resizebox{\hsize}{!}{\includegraphics{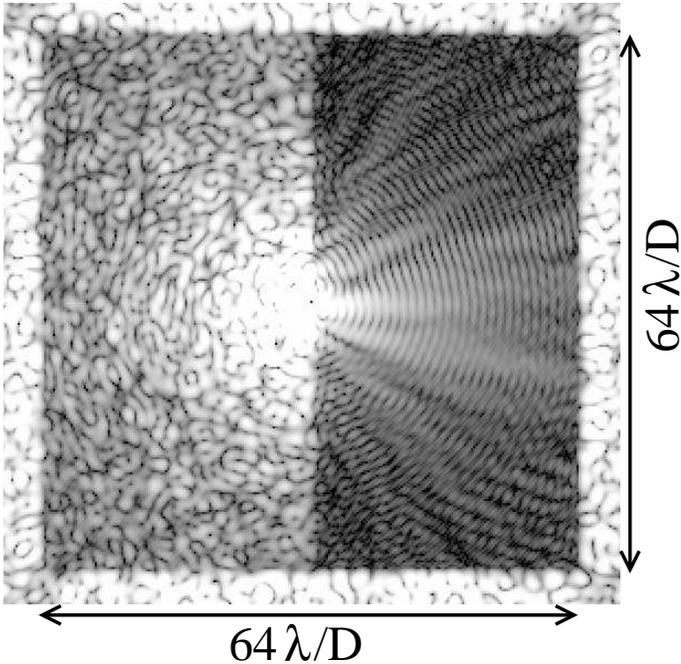}}
  \caption[Image with amplitude aberration]{\it SCC~Science image after~$4$ steps of correction. Phase errors are set to~$20$\,nm rms and amplitude aberrations to~$1\%$~rms before correction.}
  \label{fig : ampl_phase_imscc}
\end{figure}

\begin{figure}[!ht]
  \resizebox{\hsize}{!}{\includegraphics{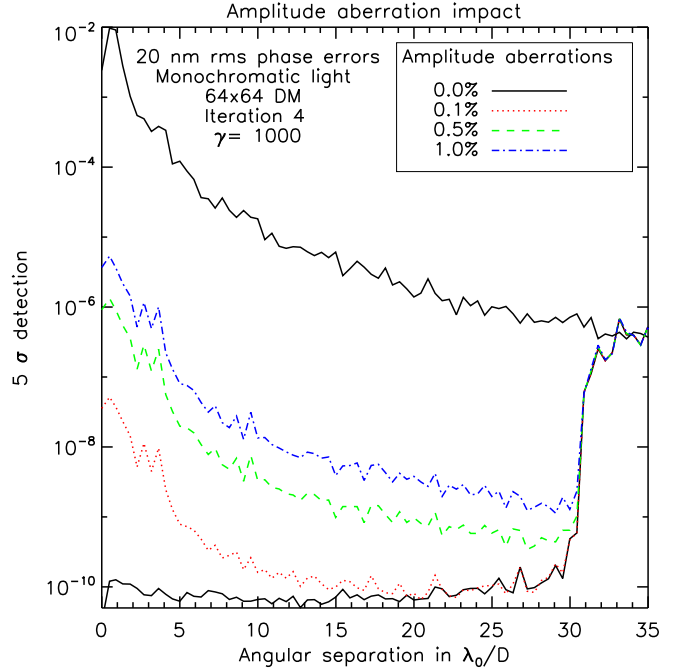}}
  \caption[Amplitude aberration impact]{\it Detection at~$5\,\sigma$~$d_{5\,\sigma}$ against the angular separation in the image at the~$4^{\mathrm{th}}$ iteration of the correcion loop. We recall the coronagraphic residue without any correction~(top black solid line).}
\label{fig : amplerr_detect}
\end{figure}

In that section, we look at the impact of amplitude errors in the pupil plane upstream of the coronagraph. We simulated the correction loop for several amplitude aberration levels~($0\%$ to $1\%$~rms). With a sole~DM, amplitude aberrations induce a reduction of the corrected area by a factor~$2$ and speckles of the uncorrected half-area diffract their light into the corrected half-area as seen in~Fig.~\ref{fig : ampl_phase_imscc} --~$1\%$ amplitude aberrations and other assumptions detailed in Table~\ref{tab : assumptions}.
In~Fig.~\ref{fig : amplerr_detect}, we plot the~$5\,\sigma$ detection~$d_{5\,\sigma}$~(Sect.~\ref{subsec : criterion}) at the~$4^{\mathrm{th}}$ iteration versus the angular separation for several amplitude aberration levels -- see assumptions in Table~\ref{tab : assumptions}.
The best correction depends roughly quadratically to the amplitude aberration level. For example, with~$1\%$~rms amplitude errors, the~$5\,\sigma$ detection is limited to~$\sim10^{-7}$ at~$\sim5\,\lambda_0/D$. It is quite important to notice that this effect is a general limitation independent from the self-coherent camera. Whatever the technique used to estimate for the phase and amplitude errors, it will be required to control with high accuracy the amplitude aberrations -- less than~$\sim1/1000$~rms seems to be a minimum with an~$f^{-3}$~PSD -- to achieve the performance limited by the~DM. As explained in Sect.~\ref{sec : DM}, using an apodized pupil might relaxes that specification. However, apodizing a pupil introduces other issues~(throughput or manufacturing limitation) and a trade-off is needed.

\section{Chromatism impact}
\label{sec : chrom}
\begin{figure}[!ht]
    \resizebox{\hsize}{!}{\includegraphics{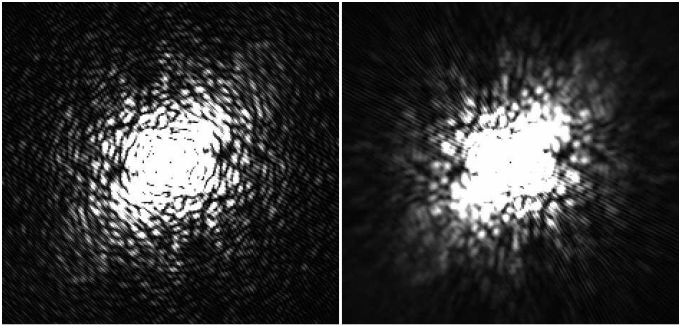}}
    \caption[SCC images in polychromatic light]{\it Science images in monochromatic~(left) and polychromatic~($R_\lambda=10$, right)  lights. These images correspond to the iteration~$0$~(no wavefront correction).}
    \label{fig : blurred}
  \end{figure}

The point spread function~(PSF) size and the interfringe are both proportional to wavelength~(cf.~Eq.~\ref{eq : intens_focal}). In white light~(small~$R_\lambda=\lambda_0/\Delta\lambda$), the science image is the superposition of all the monochromatic images over the considered bandwidth and fringes are blurred as seen in~Fig.~\ref{fig : blurred}.
As a consequence, speckles are not encoded well and cannot be corrected with accuracy. Here, we present how chromatism limits wavefront estimation and we propose a hardware solution to use the self-coherent camera in polychromatic light~($R_\lambda=\lambda_0/\Delta\lambda\simeq 5-10$). In Sect.~\ref{subsec : fieldlimit}, we study the field limitation for large bandwidths. In Sect.~\ref{subsec : factor}, we empirically optimize the chromatic factor~$F$ applied in~Eq.~\ref{eq : phase_est}. Then, in Sect.~\ref{subsec : chromlimit}, we quantify how the averaged contrast in the corrected area is modified when the bandwidth is enhanced. Finally, we  describe a hardware solution to use the~SCC in large bandpass in Sect.~\ref{subsec : wynne}.

\subsection{Field limitation}
\label{subsec : fieldlimit}
\begin{figure}[!ht]
    \resizebox{\hsize}{!}{\includegraphics{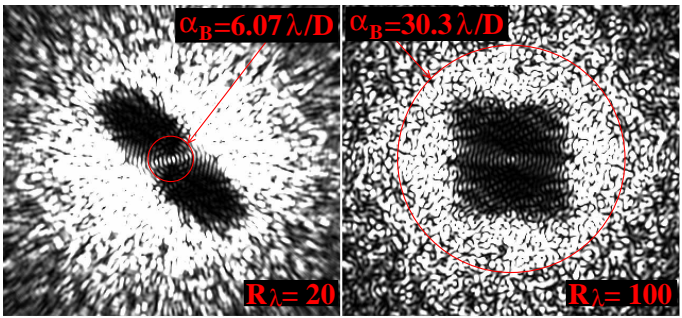}}
 \caption[Polychromatic corrected image]{Science image for the~$10$th iteration of the correction for two bandwidths: $R_\lambda=20$~(left) and~$100$~(right). We assume a~$32$x$32$ deformable mirror. The fringe direction is from bottom left to top right.}
\label{fig : fieldlimit}
 \end{figure}

We call~$\lambda_{max}$ and~$\lambda_{min}$ the largest and the smallest wavelengths of the spectral bandwidth~$R_\lambda$. The mean interfringe is~$\lambda_0/\xi_0$~(see~Eq.~\ref{eq : xi_0}). We assume the white fringe -- null OPD -- is in the center of the image. We derive the distance~$\alpha_{\mathrm{B}}$ where the fringe systems for~$\lambda_{max}$ and~$\lambda_{min}$ are shifted by half an interfringe:
\begin{equation}
\alpha_{\mathrm{B}} = \frac{\lambda_0}{\xi_0}\Big(1+\frac{1}{2\,R_\lambda}\Big)\,n = \frac{\lambda_0}{\xi_0}\Big(1-\frac{1}{2\,R_\lambda}\Big)\,n+\frac{\lambda_0}{2\,\xi_0},
\end{equation}
with~$n=\alpha_{\mathrm{B}}\,\xi_0/\lambda_{max}$. We deduce~$n=R_\lambda/2$ and for~$R_\lambda\ge 1$:
\begin{equation}
\alpha_{\mathrm{B}} \simeq \frac{R_\lambda}{2}\frac{\lambda_0}{\xi_0}.
\label{eq : alpha_b}
\end{equation}
In the perpendicular fringe direction, fringes become blurred as from~$\alpha_{\mathrm{B}}$. In the fringe direction, fringes are not blurred since~OPD between the two channels is the same for all wavelengths of the bandwidth. In that way, speckles of the science image are correctly encoded in the fringe direction whereas they are not in the perpendicular fringe direction as from~$\alpha_{\mathrm{B}}$. This effect is visible in the~Fig.~\ref{fig : fieldlimit} where we show the~$10$th iteration of the correction for a~$32$x$32$ deformable mirror and two bandwidths. For these images, we do not apply the chromatic factor~$F(\alpha)$~(Eq.~\ref{eq : F}) in the wavefront estimator~(Eq.~\ref{eq : phase_est}). Even if fringes are not blurred in the fringe direction, the speckles are dispersed~(PSF size is wavelength dependent), and wavefront errors are not estimated perfectly. Hereafter, we study the combined impacts of both effects~(fringe and speckle dispersions).
Chromatism limits the corrected area as soon as~$\alpha_\mathrm{B}$ is smaller than the~DM cutoff spatial frequency, in the worst case~$\sqrt{2}\,\rho_\mathrm{C}$ with~$\sqrt{2}$ there to account for the fringe direction). Using an~$N_\mathrm{act}$x$N_\mathrm{act}$ deformable mirror,~$\rho_\mathrm{C}=(N_\mathrm{act}/2)\,\lambda_0/D$. We deduce from~Eq.~\ref{eq : alpha_b} the minimum spectral resolution~$R_{\lambda\,\mathrm{min}}$ to have no blurred speckles in the corrected area:
\begin{equation}
R_{\lambda\,\mathrm{min}}\simeq N_\mathrm{act}\,\sqrt{2}\,\frac{\xi_0}{D}.
\end{equation}
where~$\xi_0$ is the separation between the reference and the image pupils~(cf.~Table~\ref{tab : notations}). If we consider~$\xi_0\simeq1.65\,D$~(Eq.~\ref{eq : xi_0}), we find~$R_{\lambda\,\mathrm{min}}\simeq37$,~$75$, and~$150$, respectively, for a~$16$x$16$, $32$x$32$, and~$64$x$64$ deformable mirror.

\subsection{Chromatic factor}
\label{subsec : factor}
The wavefront error estimator given by~Eq.~\ref{eq : phase_est} requires knowledge of the chromatic factor~$F$ that we optimize in the current section. In~\citet{Galicher08}, we used~$F=F_\mathrm{th}$ derived from the model of light propagation through~SCC:
\begin{equation}
  F_{\mathrm{th}}(\alpha) = \displaystyle \int_{R_\lambda}\frac{1}{\lambda^2}\exp{(2\,i\,\pi\,\alpha.\xi_0/\lambda)}d\lambda.
  \label{eq : F}
\end{equation}
However,~$F_{\mathrm{th}}$~(close to cardinal sin function) has zero values in the corrected area as soon as~$\alpha_{\mathrm{B}}\lesssim\rho_{\mathrm{C}}$~($F_{\mathrm{th}}(\alpha_{\mathrm{B}})=0$). Speckles located at the zeros of~$F_{\mathrm{th}}$ are not encoded well and can be neither estimated nor corrected. This limitation is similar to the zero values of the reference amplitude~(Sect.~\ref{subsec : gamma}). Using the estimator of~Eq.~\ref{eq : phase_est}, pupil spatial frequencies corresponding to separations higher than~$\alpha_{\mathrm{B}}$~(blurred speckles) cannot be recovered. But we can restrain their impact on the corrected area, and we test five chromatic factors:
\begin{eqnarray}
 F_1 &=&F_\mathrm{th}  \nonumber\\
 F_2&=&\frac{1}{F_\mathrm{th}} \\
 F_3&=&\frac{1}{F^*_\mathrm{th}} \nonumber \\
 F_4&=&1 \nonumber\\
 F_5&=&1 \qquad\mathrm{if}\qquad (F_\mathrm{th}+F^*_\mathrm{th})\ge 0  \nonumber
\end{eqnarray}
We attempt to correct for unblurred speckles~(Sect.~\ref{subsec : fieldlimit}) and minimize the impact of the uncorrected speckles~(blurred ones). To compare the different chromatic factors, we consider assumptions given in Table~\ref{tab : assumptions} and look at the evolution of the averaged contrast in the corrected area~$C_1$ during the correction for~$R_\lambda=40$. We find that setting the factor~$F$ to~$1/F_{\mathrm{th}}$ provides the best results~(no divergence). That factor reduces the impact of the uncorrected speckles because it multiplies their intensity by a low value in the estimator of~Eq.~\ref{eq : phase_est}. It acts as a regularization of the uncorrected spatial frequencies. We use that chromatic factor in all the paper from now. A more sophisticated estimator~(regularization of a $\chi^2$ minimization without linearization of wavefront errors and without assumptions on~$R_\lambda$) could certainly be developed but has not been studied yet.

\subsection{Correction level}
\label{subsec : chromlimit}
\begin{figure}[!ht]
  \resizebox{\hsize}{!}{\includegraphics{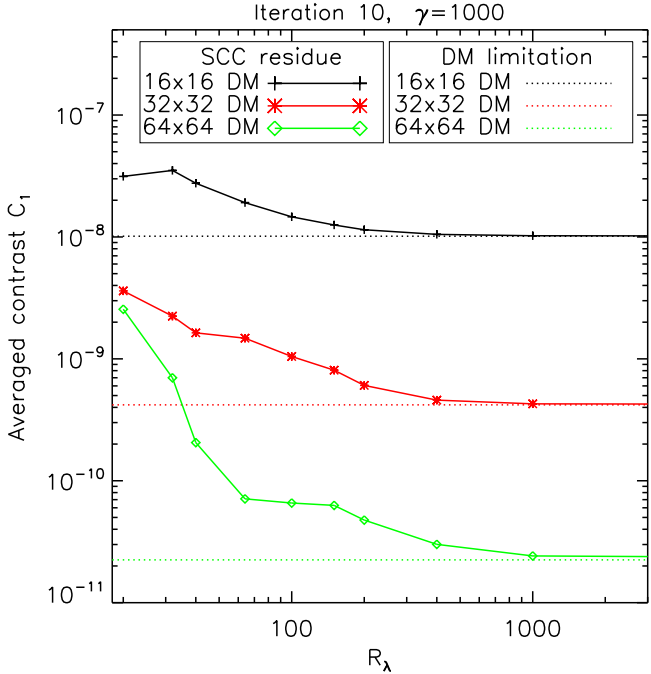}}
  \caption[Chromatism study]{\it Averaged contrast~$C_1$ in the corrected area at the~$10$th iteration of the correction versus the spectral resolution~$R_\lambda$ for three deformable mirrors.}
  \label{fig : chrom}
\end{figure}

The convergence speed of the correction loop slightly decreases when the light becomes more and more chromatic: $3$~steps in monochromatic light and~$\sim10$~steps for~$R_\lambda\simeq30$ for a~$32$x$32$. At the same time, the averaged correction gets worst as seen in~Fig.~\ref{fig : chrom} where the criterion~$C_1$ at the~$10th$ loop iteration is plotted against the spectral bandwidth~($R_\lambda=20$ to~$\infty$) under the assumptions of Table~\ref{tab : assumptions}.
In quasi-monochromatic light, the DM limitation is achieved~(cf.~Sect.~\ref{sec : DM}) for all the DM sizes. Naturally, the correction is more efficient when the number of actuators increases. The minimum~$R_\lambda$ required to have the correction converging to less than twice the~DM limitation in fewer than~$5$ steps increases with the size of the corrected area: $\sim80$ for the~$16$x$16$, $\sim100$ for the~$32$x$32$, and~$\sim200$ for the $64$x$64$. These values are close to the expected ones determined in Sect.~\ref{subsec : fieldlimit}.

From these results, it is clear that chromatism strongly limits~SCC performance and that the self-coherent camera cannot be used with a classical bandwidth~$R_\lambda\simeq 5$. Working with a narrow bandwidth is not the solution because of photon noise. We foresee two possibilities. We can develop software solutions and modify the wavefront estimator of~Eq.~\ref{eq : phase_est} to account for polychromatic dispersions of speckles and fringes~(regularization of a~$\chi^2$ minimization without linearization of aberrations and without assumptions on~$R_\lambda$). Hardware solutions are also conceivable. We could associate the self-coherent camera with an integral field spectrometer~(IFS) at modest resolution~($R_\lambda\sim 100$). We could estimate wavefront errors for each wavelength channel. We could also develop a new algorithm to process the data of all the channels at the same time to optimize the estimation. This solution will be studied in future work. Another hardware solution is a Wynne compensator that we describe in Sect.~\ref{subsec : wynne}.

\subsection{Wynne compensator}
\label{subsec : wynne}
\begin{figure}[!ht]
    \resizebox{\hsize}{!}{\includegraphics{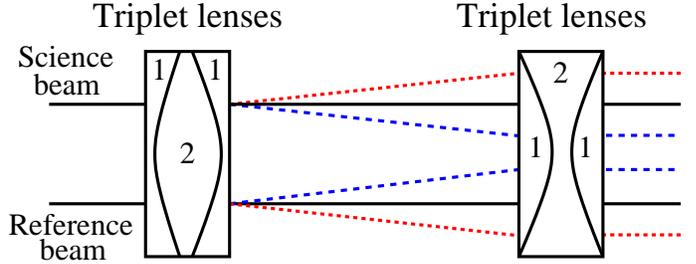}}
    \caption[Wynne compensator scheme]{\it Scheme of a Wynne Compensator composed by two triplet lenses made with two kinds of glasses~(called~$1$ and $2$). Dispersion is linear with the wavelength and the outgoing beam is collimated.}
    \label{fig : wynne_schema}
  \end{figure}

The image widths~(about $\lambda/D$ and $\lambda/D_{\mathrm{R}}$) and the fringe period~($\lambda/\xi_0$) are proportional to wavelength, which is the reason for the~SCC chromatism limitation~(Sects.~\ref{subsec : fieldlimit} to~\ref{subsec : chromlimit}). In the context of speckle interferometry~\citep{labeyrie70}, \citet{Wynne79} proposed a device to correct for such a spectral dependence over a wide spectral range: the Wynne compensator~(Fig.~\ref{fig : wynne_schema}).
 This device is composed of two triplets of lenses. In each triplet, the first and the third lenses are made in the same glass, called glass~$1$, while the second lens glass is different, glass~$2$. Each second lens is symetric~(same curvature on each side). Indexes of glasses~$1$ and~$2$ are equal at the middle wavelength of the considered bandpass, but their dispersions are different. The beam is then not modified for the middle wavelength~(black solid curve), while it is magnified for other wavelengths~(red dotted and blue dashed curves). The compensator's outgoing beam is collimated for all wavelengths. In~Fig.~\ref{fig : wynne_schema}, we only show the average beam of image and reference channels for reasons of clarity. In that way, we only outline that the Wynne compensator magnifies the separation~$\xi_0(\lambda)$ between pupils. A more detailed study of the Wynne compensator shows it also magnifies pupil diameters~$D(\lambda)$ and~$D_{\mathrm{R}}(\lambda)$. By choosing the right glasses and optimizing the curvatures of the lenses, their thicknesses and the distance between the two triplets, we can apply a magnification proportional to the wavelength over a large bandwidth in visible light.

\begin{figure}[!ht]
    \resizebox{\hsize}{!}{\includegraphics{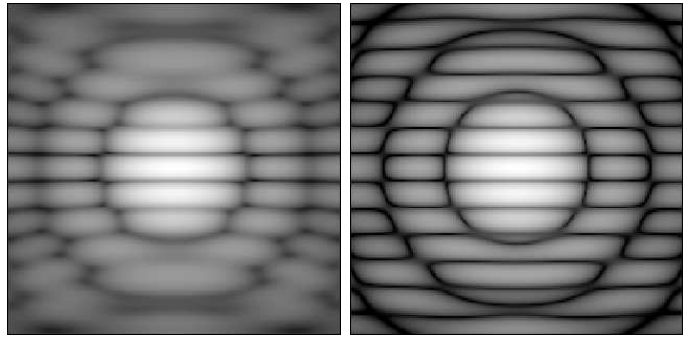}}
    \caption[Wynne compensator images]{\it Science image without~(left) or with~(right) Wynne compensator. The initial spectral resolution is~$6.5$ at~$650\,nm$. For the corrected image, the effective bandwidth is about~$0.75\%$~(right). We use log scale.}
    \label{fig : interf_wynne}
  \end{figure}

Such a Wynne compensator can be associated to the~SCC by adding it just before the recombining optic. Since magnification is proportional to the wavelength,~$D$,~$D_{\mathrm{R}}$, and~$\xi_0$ are proportional to~$\lambda$ and we obtain non-blurred fringes all over the detector~(corresponding to the monochromatic case). A simulation is shown in~Fig.~\ref{fig : interf_wynne} for a spectral resolution of~$R_\lambda=6.5$ at~$650$\,nm. Since only magnification is important, we do not consider any coronagraph and the incoming wavefront is assumed to be aberration-free.
The relative difference between the Airy pattern widths and interfringes for the extreme wavelengths~($600$ and $700$\,nm) is about~$0.75\%$, which corresponds to~$R_\lambda\simeq130$. That spectral resolution~($130$) is the spectral bandwidth required to have almost no blurred speckles in the corrected area of a~$64$x$64$ deformable mirror~(Sects.~\ref{subsec : fieldlimit} and~\ref{subsec : chromlimit}). An optimized Wynne compensator would enable us to use the self-coherent camera in polychromatic light~($R_\lambda=6.5$ in visible light~$\lambda_0=650$\,nm) and achieve the quasi-monochromatic performance~(Fig.~\ref{fig : chrom}).

Finally, even if chromatism seems to be a hard point of the self-coherent camera technique, several solutions are conceivable: more sophisticated estimators, hardware solutions~(association with a Wynne compensator or an integral field spectrometer at modest resolution).

\section{SCC and real coronagraphs}
\label{sec : sccrealcoro}
The previous sections present a parametric study of the self-coherent camera without mentioning any concrete setup and assuming a perfect coronagraph as in~\citet{Galicher08}. In~\citet{Baudoz06} and~\citet{Galicher07}, we proposed a self-coherent camera device built as an interferometer: a beamsplitter to create the reference channel, a pinhole to filter the reference beam, a delay line to ensure a null optical path difference, and a lens to recombine image and reference beams. The disadvantage of that device is the delay line that has to be controlled with very high accuracy in real time~(Sect.~\ref{subsec : opd}). In Sects.~\ref{subsec : scclyotstop} and~\ref{subsec : phase_est_real}, we propose a new robust design for associating the self-coherent camera with a coronagraph that has a Lyot stop plane. We describe in detail the case of a FQPM~\citep{Rouan00} in Sect.~\ref{subsec : sccfqpm}.

\subsection{How to use the Lyot stop plane}
\label{subsec : scclyotstop}
\begin{figure}[!ht]
  \resizebox{\hsize}{!}{\includegraphics{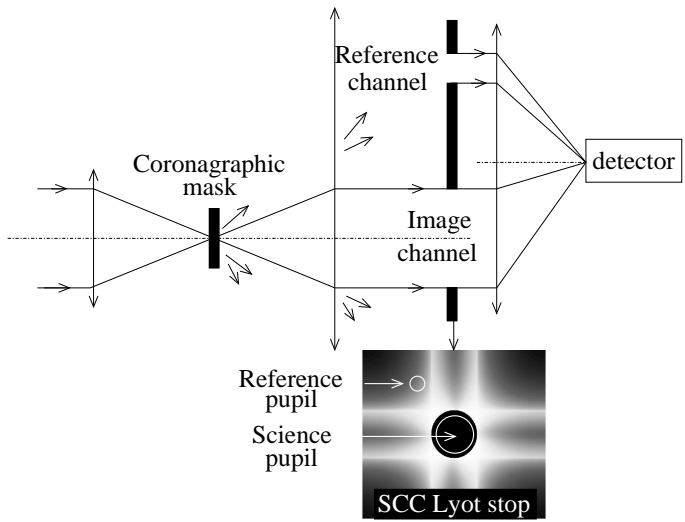}}
 \caption[]{\it Schematics of the self-coherent camera associated to a coronagraph that uses a Lyot stop plane.  The pupil intensity distribution in the Lyot stop plane is given for a~FQPM coronagraph.}
    \label{fig : sccfqpm}
 \end{figure}

Using a coronagraph that needs a Lyot stop is very interesting because it can be easily associated with the SCC. Such a coronagraph rejects only the stellar light outside the pupil -- no companion light -- so that the light stopped by the Lyot stop comes only from the hosting star and can be used to create the reference channel. We propose adding a small non-centered pupil to the classical Lyot stop as shown in the schematics of~Fig.~\ref{fig : sccfqpm} for a~FQPM coronagraph.
The sole difference from the classical coronagraphic device is the modification of the Lyot stop. That solution is very attractive because it constrains a constant optical path difference between the reference and the image channels. As explained in Sect.~\ref{subsec : wynne}, we may want to associate the SCC with a chromatic Wynne compensator~\citep{Wynne79}. This one would be placed between the SCC Lyot stop and the recombining optic.

\subsection{Wavefront estimation}
\label{subsec : phase_est_real}
The complex amplitude estimated by~Eq.~\ref{eq : psiS_0} is the pupil amplitude in the Lyot stop plane downstream the coronagraph. The second step is determine the complex amplitude~$\psi_{\mathrm{S}}'$ upstream of the coronagraph where the deformable mirror stands. If the coronagraph is perfect and wavefront aberrations are small, we can use the estimator of~Eq~\ref{eq : phase_est}. If the coronagraph is not perfect, we have to model the light propagation through it.

Calling $M$ the mask function in the focal plane and~$L$ the classical Lyot stop~(sole image channel), we have
\begin{equation}
  \psi_{\mathrm{S}} = \left(\psi_{\mathrm{S}}' *\mathcal{F}^{-1}(M)\right)\,L,
  \label{eq : FQPM}
\end{equation}
where~$*$ denotes the convolution. We deduce~$\psi_{\mathrm{S}}'$ within the~$L$ area:
\begin{equation}
  \psi_{\mathrm{S}}' = \mathcal{F}^{-1}\left[\mathcal{F}(\psi_{\mathrm{S}})\,\Big(\frac{1}{M}\Big)_0\right],
  \label{eq : psi}
\end{equation}
where~$(1/M)_0$ is the inverse of the mask function~$M$ where~$M$ is not zero and equals~$0$ elsewhere. This expresses that we cannot estimate the spatial frequencies for which the mask~$M$ has stopped the energy~(ie.~$M=0$). We finally estimate the complex amplitude upstream of the coronagraph from~Eqs.~\ref{eq : psiS_0} and~\ref{eq : psi}. We notice that phase masks are not bounded by this limitation since they do not block light.

\subsection{SCC and four-quadrant phase mask}
\label{subsec : sccfqpm}
The~FQPM coronagraph uses a Lyot stop and can be associated with the self-coherent camera in a device, called~SCC-FQPM. The~FQPM has been described in detail in previous papers~\citep{Rouan00,Riaud01,Riaud03,Boccaletti04}. We recall the coronagraphic mask~$M$ induces a~$\pi$ phase shift on two quadrants of a diagonal and no phase shift in the two other quadrants. In~Fig.~\ref{fig : sccfqpm}, we explain where the image and reference pupils are picked in the Lyot stop of an~SCC-FQPM. In Sect.~\ref{para : ref_flux_fqpm}, we first give the quantity of energy that goes through the reference pupil to determine the optimized ratio~$\gamma=D/D_{\mathrm{R}}$ of pupil diameters~(cf.~Sect.~\ref{subsec : gamma}). We then study the optical path difference between reference and image channels~(cf.~Sect.~\ref{subsec : opd}) in Sect.~\ref{para : fqpm_opd}. Finally, we examine the performance of the~SCC-FQPM under realistic conditions in Sect.~\ref{subsubsec : sccfqpm_speed}. In the whole section, we assume a perfect achromatic FQPM: infinitely thin transitions and perfect phase shifts.

\subsubsection{Implementation}
\paragraph{Reference flux\\}
\label{para : ref_flux_fqpm}
As explained in Sect.~\ref{subsec : gamma}, the ratio between image and reference pupil diameters~$\gamma=D/D_{\mathrm{R}}$ has to be optimized for each~SCC design. In Table~\ref{tab : reffluxfqpm}, we give the averaged intensities of the image and reference~PSFs in the corrected area of a~$32$x$32$~DM for different~$\gamma$ and phase-error levels. All fluxes are expressed in photons per pixel. To establish the table, we consider the device of figure~\ref{fig : sccfqpm} working under assumptions of Table~\ref{tab : assumptions}. We put the reference pupil behind one of the quadrants of the~FQPM~(see~Sect.~\ref{para : fqpm_opd}), and we set the stellar flux using assumptions of Sect.~\ref{subsec : assumptions}, a~$20$\,s exposure time, and~$4$\,m diameter telescope.
\begin{table}[!ht]
  \renewcommand{\arraystretch}{1.3}
  \begin{center}
   \begin{tabular}{|c||c|c||c|c|}
      \hline
      $\gamma$& \multicolumn{2}{|c||}{$10$} & \multicolumn{2}{|c|}{$20$} \\
      \hline
      $\sigma[\phi]$~(nm) &  $20$ & $40$ &  $20$ & $40$\\
      \hline
      $<I_{\mathrm{R}}>$ & $212$ & $193$& $58$ & $53$ \\
      \hline
      $<I_{\mathrm{S}}>$ & $45835$ & $176218$& $45835$ & $176218$\\
      \hline
      $\frac{<I_{\mathrm{R}}>}{<I_{\mathrm{S}}>}\,10^{2}$ & $0.46$ & $0.11$& $0.13$ & $0.03$\\
      \hline
    \end{tabular}
   \caption{\it Averaged fluxes of image~($<I_{\mathrm{S}}>$) and reference~($<I_{\mathrm{R}}>$) channels in the corrected area of a~$32$x$32$~DM for different~$\gamma$~(first raw) and phase error levels~(second raw). We consider an achromatic~FQPM. Fluxes are given in photons per pixel. See Table~\ref{tab : assumptions} for more informations on assumptions.}  \label{tab : reffluxfqpm}
  \end{center}
\end{table}
Coronagraphic residue~$<I_{\mathrm{S}}>$ in the image channel depends roughly quadratically on the phase error level as expected because the coronagraphic pupil is approximately~$\psi_{\mathrm{S}}\simeq (a+i\,\phi)\,P$. In contrast, the rejected star energy remains slightly constant in the reference pupil~($<I_{\mathrm{R}}>$). We notice also that~$<I_{\mathrm{R}}>$ roughly evolves as the inverse square of~$\gamma$. We propose setting~$\gamma$ to~$\sim20$ so that the condition of a few photons per detector pixel coming in from the reference channel~(see Sect.~\ref{subsec : irmin}) can be respected with exposure times that are not too long~(speckle lifetime). It is important to notice that this choice is not strict. To strictly set~$\gamma$, we have to know the lifetime of the speckles to be corrected. This would impose a maximum exposure time to record the interferential images~$I$~(aberrations have to be static to be estimated from~$I$). From that maximum time, we could choose the apropriate~$\gamma$ value to let the reference flux verify the condition of Sect.~\ref{subsec : irmin}. To make that choice, we may notice again that higher values of~$\gamma$ are interesting to ensure stability of the reference image and to limit the presence of low values of~$A_{\mathrm{R}}$ in the corrected area~(cf.~Sect.~\ref{subsec : gamma}).

\paragraph{Optical path difference\\}
\label{para : fqpm_opd}
We have shown in Sect.~\ref{subsec : opd} that we need to accurately control the~OPD between reference and image channels~(accuracy and stability of~$\sim\lambda_0/6$). In the case of an~SCC-FQPM, if the reference pupil is picked behind one of the~$\pi$ phase shift quadrants of the focal mask, the optical path difference in the Lyot stop plane in monochromatic light is~$\pi$. Fringes are shifted by half an interfringe, and we have to multiply by~$-1$ the estimator of~Eq.~\ref{eq : psi}. If the reference pupil is picked behind another quadrant, the~OPD is null and we can use directly the estimator of~Eq.~\ref{eq : psi}. The directions to avoid are the transition direction of the FQPM for which the~OPD is not just~$0$ or~$\pi$ and above all is not uniform over the reference pupil so that it is not easy to account for it in the estimator. In a real experiment, we propose to pick the reference pupil behind one of the four quadrants avoiding the transition directions and to calibrate the sign of the estimator. We are confident that the~OPD is stable in time in the compact device proposed in~Fig.~\ref{fig : sccfqpm} and laboratory demonstrations are in progress to confirm it.

\subsubsection{Performances}
\label{subsubsec : sccfqpm_speed}
\begin{figure}[!ht]
  \resizebox{\hsize}{!}{\includegraphics{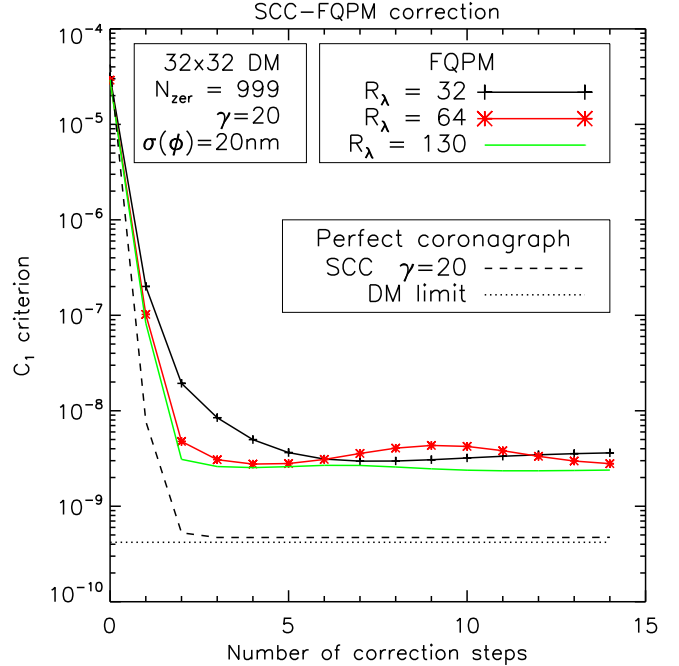}}
 \caption[$C_1$ convergence for a SCC-FQPM]{\it Averaged contrast~$C_1$ in the corrected area of a~SCC-FQPM versus the number of correction iterations for several bandwidths~$R_\lambda=\lambda_0/\Delta\lambda$. The dot-and-dash line and the dashed line represent respectively the best~DM correction and the monochromatic case with a perfect coronagraph.}
\label{fig : figchromfqpm}
\end{figure}

\paragraph{Impact of chromatism\\}
\label{para : sccfqpm_chrom}
If we directly apply the wavefront estimator of~Eq.~\ref{eq : psi}, the correction loop becomes slower than in the perfect coronagraph case~(Sect.~\ref{subsec : chromlimit}) because the~FQPM model is not perfect. We then employ a more complex model accounting for the exact~FQPM impact on the first~$999$ Zernike polynomials~\citep{Noll76}. Using that model, we plot the averaged contrast~$C_1$ in the corrected area of a~SCC-FQPM versus the correction step for several spectral bandwidths~(Fig.~\ref{fig : figchromfqpm}). Assumptions are given in Sect.~\ref{subsec : assumptions} and Table~\ref{tab : assumptions}.
The same remarks as in a perfect coronagraph case~(Sect.~\ref{subsec : chromlimit}) are valid: convergence slows down and gets worst when the bandwidth gets wider~($R_\lambda$ decreases). However, the SCC-FQPM best correction is around five times less effective than with a perfect coronagraph~(dashed line) which reaches the~DM limitation~(dot-and-dash line, cf.~Fig.~\ref{fig : dmlimitsize}). This is because the~FQPM does not well estimate all optical defects. For example, the astigmatism in the FQPM transition direction is very poorly estimated~(study not shown in this paper). We work on that limitation to overcome it, and we also plan to use an other coronagraph mask without such effects as the annular groove phase mask~\citep{Mawet05}. However, even with that limitation, the correction is effective and SCC-FQPM provides very high-contrast imaging as shown in the next section.

\begin{figure}[!ht]
  \resizebox{\hsize}{!}{\includegraphics{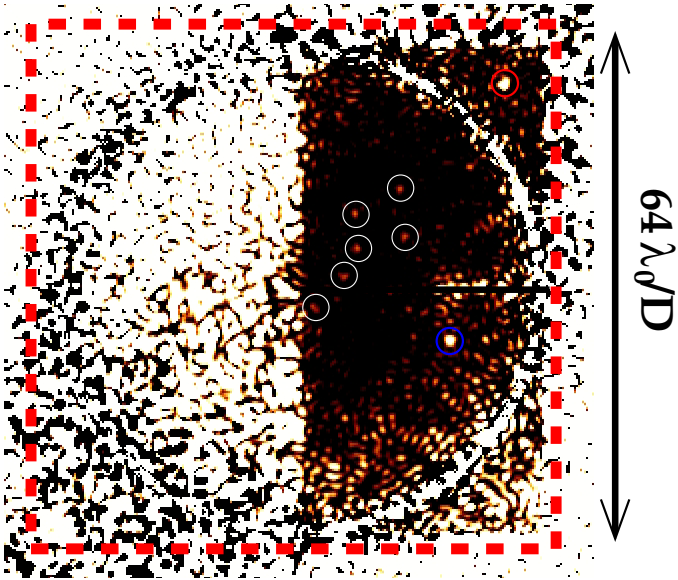}}
  \caption[Earth, super-earth and Jupiter detections with a SCC-FQPM under realistic conditions]{\it Detections of earths~(white circles), super-earth~(blue circle) and Jupiter~(red circle) with a SCC-FQPM downstream a~$4$\,m space telescope and a~$64$x$64$ deformable mirror. The image is the result of a wavefront correction loop~(step~A, $7$\,h$25$\,min) and a companion estimation~(step~B).}
\label{fig : sccfqpm_detection}
\end{figure}

\paragraph{SCC-FQPM detections\\}
\label{para : sccfqpm_detect}
In the last section, we check for the detection efficiency of SCC-FQPM under realistic assumptions detailed in Sect.~\ref{subsec : assumptions} and~Table~\ref{tab : assumptions} for spatial observations. The variance of the whole phase error is~$20$\,nm~rms, and we set the initial astigmatism defects in the~FQPM transition direction to~$1$\,nm~rms~(levels before correction). We account for amplitude aberrations of~$1\%$~rms. We simulated twelve~$2\,10^{-10}$ earths, located symmetrically with respect to the image center at~$3$, $6$, $9$, $12$, $15$, and $18\,\lambda_0/D$. For the~$4$\,m telescope that we consider~(throughput of~$40\%$), these separations correspond to~$0.12$~($1.24$), $0.25$~($2.48$), $0.37$~($3.71$), $0.50$~($4.95$), $0.62$~($6.19$), and $0.74$~($7.43$)\,arcsec~(AU). One~$8\,10^{-10}$ super-earth is located at~$20\,\lambda_0/D$~($8.25$\,AU) and a~$10^{-9}$ Jupiter-like planet sits at~$36.5\,\lambda_0/D$~($15$\,AU). All spectra are flat and the hosting star is a~G$2$ star at~$10$\,pc. We closed the loop to correct for wavefront errors~(step~A, Sect.~\ref{subsec : stepa}) during~$7$\,h$25$\,min, which corresponds to~$6$ iterations. After step~A, we apply the companion estimator and obtain the image presented in~Fig.~\ref{fig : sccfqpm_detection}. We consider photon noise but no read-out noise.
The white circle of uncorrected speckles corresponds to the low values of the reference amplitude for~$\gamma=25$. Stellar speckles at these positions are not correctly encoded~(estimated) and they are not corrected~(Sect.~\ref{sec : snr}). The red dashed box bounds the centered~$64$x$64\,(\lambda_0/D)^2$ area. Five earths are clearly detected~(white thin circles) in the first quadrant at~$0.25$~($6$), $0.37$~($9$), $0.50$~($12$), $0.62$~($15$), and~$0.74$~($18$)\,arcsec~($\lambda_0/D$). The sixth at~$3\,\lambda_0/D$ is detected~(white thin circle in the fourth quadrant), but speckle noise is just below its level. Positions of the six other earths are symmetric with respect to the image center and are in the uncorrected area~(amplitude aberrations limitation). Intensities of the super-earth in the fourth quadrant~(blue circle) and of the Jupiter on the top right~(red circle beyond the reference ring) are well above the surrounded speckle field. Table~\ref{tab : contrast_sccfqpm} gives the measured flux and position for each detected planet. Fluxes are slightly underestimated~(around~$10\%$ accuracy). Positions are accurate at~$\sim0.25\,\lambda_0/D$, which equals to~$10$\,mas~($0,10$\,AU for a~$10$\,parsec star).
\begin{table*}[!ht]
  \renewcommand{\arraystretch}{1.2}
  \renewcommand{\multirowsetup}{\centering}
  \begin{center}
      \begin{tabular}{||c|c||c|c||}
     \hline
     \hline
\multicolumn{2}{||c||}{Contrast~($\times 10^{10}$)}&\multicolumn{2}{|c||}{Angular separation in~$\lambda_0/D$}\\
\hline
Simulated value & Measured value & Simulated value & Measured value\\
   \hline     
\multirow{6}{*}{$2$} & $1,7$& $3,0$& $3,0$\\
\cline{2-4}
 & $1,5$& $6,0$& $5,9$\\
\cline{2-4}
 & $1,8$& $9,0$& $8,9$\\
\cline{2-4}
 & $2,1$& $12,0$& $11,8$\\
\cline{2-4}
 & $1,6$& $15,0$& $14,7$\\
\cline{2-4}
 & $1,6$& $18,0$& $18,0$\\
     \hline     
$8$ & $7,7$& $20,0$& $19,9$\\
     \hline
$10$ & $9,1$& $36,5$& $36,3$\\
     \hline
     \hline
    \end{tabular}
    \caption[Measured contrasts and angular separations]{\it Comparison between the contrasts and angular separations measured in the~SCC image and the simulated values.}
\label{tab : contrast_sccfqpm}
\end{center}
\end{table*}

Finally, planets as faint as earths are detectable by~SCC-FQPM in a few hours from space under realistic assumptions.

\section{Conclusion}
In Sect.~\ref{sec : DM}, we provided the intrinsic limitation for deformable mirrors controlled by the algorithm of~\citet{Borde06}, under realistic yet optimistic assumptions~(no dead actuators, continuous face sheet). It is important to keep in mind that this limitation does not depend on the technique used to estimate for wavefront errors since we assumed a perfect estimation. One way to improve the~DM best contrast could be an apodization of the pupil so that the uncorrected speckles would diffract their light in a more restricted area. However, all the techniques proposed to apodize a pupil~\citep{vanderbei03,kasdin05,Guyon05,Pluzhnik06} come with throughput problems or manufacturing limitations.

In Sects.~\ref{sec : snr} to~\ref{sec : chrom}, we gave the results of the parametric study of a self-coherent camera~(SCC) used as a focal plane wavefront sensor and associated with a perfect coronagraph and a deformable mirror. Several points do not seem to be critical for the technique: reference flux~(Sect.~\ref{sec : snr}), error on the exact position of the reference image, and diameter of the reference pupil~(Sect.~\ref{subsec : gamma}). On the contrary, two points are more critical:
\begin{itemize}
\item {\bf optical path difference between the reference and the image channels.} We have to know and control this~OPD with an accuracy of about~$\lambda_0/6$. If we associate the SCC with a coronagraph using a Lyot stop as described in Sect.~\ref{sec : sccrealcoro}, the hardwar optical path difference is always zero and we only have to control the end of the setup. Only common optics are used in this setup. We could put the device in a closed box to avoid differential air variations. We plan to check for the level of the variations in the optical path difference in the device of~Fig.~\ref{fig : sccfqpm} in a laboratory experiment.
\item {\bf chromatism.} As shown in Sect.~\ref{sec : chrom}, chromatism is the most critical point in the technique. The main consequence is the reduction of the corrected area, in other words, the field of view of the image as shown in~Fig.~\ref{fig : fieldlimit}. The uncorrected speckles spread light and limit the contrast of the detection. We are studying software and hardware solutions to minimize that effect. For example, for the former we could develop a more sophisticated wavefront estimator using a~$\chi^2$ minimization with regularization terms to account for the spectral dispersion of the speckles~($\lambda_0/D$) and of the fringes~($\xi_0/D$). Hardware solutions are certainly more appropriate. We presented the Wynne compensator in Sect.~\ref{subsec : wynne}. According to numerical simulations, it would enable working with a classical bandpass~($\sim15\%$, $R_\lambda\simeq6$) in the visible light with a~$64$x$64$~DM. We will test such a Wynne compensator in a laboratory experiment. A second hardware solution to overcome the chromatism limitation could be the association of the~SCC with an integral field spectrometer at modest spectral resolution~($R_\lambda\simeq100-150$). This solution is very attractive because it would directly provide a companion spectra, but no work has been done on it yet.
\end{itemize}

We showed in Sect.~\ref{sec : amplitude} that the self-coherent camera can estimate for both phase and amplitude aberrations. The impact of the amplitude aberrations is quite critical: their level must be smaller than~$1/1000$ to reach a~$5\,\sigma$ detection of~$10^{-9}$ at~$5\,\lambda_0/D$ with an~$f^{-3}$ power spectral density. That limitation is not intrinsic to the~SCC and would limit any high-contrast imaging system.

Section~\ref{sec : sccrealcoro} presented a very simple and robust design that associates the self-coherent camera with any coronagraph which uses a Lyot stop. The most interesting point is that the optical path difference between the two channels is constant per construction. In Sect.~\ref{subsec : sccfqpm}, we studied in detail the case of the association of the~SCC with a~FQPM coronagraph~(SCC-FQPM) and we showed in Sect.~\ref{subsubsec : sccfqpm_speed} that the performance is very attractive and comparable to the case of a perfect coronagraph that is presented in~\citet{Galicher08}. Detections of~$2\,10^{-10}$ earths, $8\,10^{-10}$ super-earth, and~$10^{-9}$ Jupiter under realistic assumptions are numerically demonstrated for an~SCC-FQPM in polychromatic light~($R_\lambda=5$) using a Wynne compensator~(reducing the effective bandwidth to~$R_{\lambda\,\mathrm{eff}}=150$) in $\sim7$\,h$25$\,min with a~$4$\,m space telescope.

The next steps are laboratory demonstrations of both~SCC capabilities: focal-plane wavefront estimation~(step~A, Sect.~\ref{subsec : stepa}) and companion estimation by differential imaging~(step~B, Sect.~\ref{subsec : stepb}). We will also attempt to overcome the poor estimate of the astigmatism in the~FQPM transition direction. New algorithms have already been developed for using a~DM interaction matrix including the impact of the whole instrument: DM, coronagraph, and~SCC. It will be tested in a laboratory experiment very soon. We also study the~SCC association with other coronagraphs like an annular groove phase mask~\citep{Mawet05}.

\vspace{1cm}
We thank R\'{e}mi~Soummer for private communications about his paper ``Fast computation of Lyot-style coronagraph propagation''\citep{Soummer07b}, which was very useful for simulating the polychromatic images and the different values of the~$\gamma$ parameter.

\section*{Appendix A}
\label{ap : apA}
In that appendix, we present how photon noise and read-out noise propagate through the numerical algorithms providing the wavefront estimation~(Eq.~\ref{eq : phase_est}) and the companion estimation~(Sect.~\ref{subsec : icmin}). We call~$I_n(\alpha)$ the noisy intensity of the recorded sience image:
\begin{equation}
I_n(\alpha) = I(\alpha)+\epsilon(\alpha),
\label{eq : intens_noise}
\end{equation}
where~$I(\alpha)$ is the noiseless intensity at position~$\alpha$ of the interferential image~(Eq.~\ref{eq : intens_focal}) and~$\epsilon(\alpha)$ the noise. For photon noise~(Poissonian distribution), the mean of~$\epsilon$ is zero and its variance~$Var[\epsilon(\alpha)]$ at the position~$\alpha$ is~$I(\alpha)$. For a uniform read-out noise~(gaussian distribution), the mean of~$\epsilon$ is null and  its variance is~$\sigma^2_{\mathrm{RON}}$ over the whole image. For both photon noise and read-out noise, we assume the noise is not correlated between pixels located at different positions~$\alpha_1$ and~$\alpha_2$~(the spatial covariance of~$\epsilon$ is zero):
\begin{equation}
E[\epsilon(\alpha_1)\,\epsilon^*(\alpha_2)] = Var[\epsilon(\alpha_1)]\,\delta(\alpha_1-\alpha_2),
\label{eq : covar_eps}
\end{equation}
where~$\delta$ is the Kronecker symbol and~$E$ the statistical mean. The self-coherent camera technique uses~$\mathcal{F}^{-1}[I_n]$, the inverse Fourier transform of the science image~$I_n$. Let~$\nu(u)$ be the noise on~$\mathcal{F}^{-1}[I_n](u)$ at the position~$u$ in the Fourier plane, such that:
\begin{equation}
\mathcal{F}^{-1}[I_n](u) = \mathcal{F}^{-1}[I](u)  + \nu(u).
\label{eq : ft_intens_noise}
\end{equation}
Using~Eqs.~\ref{eq : intens_noise} and~\ref{eq : ft_intens_noise}, the monochromatic case at~$\lambda_0$ and the Fourier transform properties, we find that the mean of~$\nu$ is zero and its spatial covariance is
\begin{eqnarray}
  & &E[\nu(u_1)\,\nu^*(u_2)] =  \nonumber \\
&    & \iint{\exp{\left[\frac{2\,i\,\pi\,(\alpha_1.u_1-\alpha_2.u_2)}{\lambda_0}\right]}\,E[\epsilon(\alpha_1)\,\epsilon(\alpha_2)]\,\mathrm{d}\alpha_1\,\mathrm{d}\alpha_2}.
\label{eq : covar_nu}
\end{eqnarray}
From~Eq.~\ref{eq : covar_eps}, we determine the covariance of~$\nu$ in the case of photon noise
\begin{equation}
E[\nu(u_1)\,\nu^*(u_2)] = \mathcal{F}^{-1}[I](u_1-u_2),
\label{eq : covar_nu_BP}
\end{equation}
and in the case of read out noise:
\begin{equation}
E[\nu(u_1)\,\nu^*(u_2)] = \sigma^2_{\mathrm{RON}}\,\delta(u_1-u_2).
\label{eq : covar_nu_BL}
\end{equation}
The spatial covariance of the inverse Fourier transform of photon noise~(Poissonian distribution) is not reduced to its variance whereas it is for read-out noise~(Gaussian distribution).

During the~SCC image processing, to estimate~$A_{\mathrm{S}}$, we isolate one of the two lateral peaks of~$\mathcal{F}^{-1}[I_n]$~\citep{Baudoz06,Galicher07} using a circular binary mask of diameter~$D+D_{\mathrm{R}}\simeq D$~(see Sect.~\ref{subsec : gamma} for that approximation) and we apply a Fourier transform. The noiseless part of that Fourier transform is called~$I_-(\alpha)$~(Eq.~\ref{eq : i_-def}). The noisy part is~$I_{-\,n}(\alpha)$
\begin{equation}
I_{-\,n}(\alpha)=I_-(\alpha)+\epsilon_-(\alpha),
\end{equation}
with~$\epsilon_-(\alpha)$ the noise at the~$\alpha$ position. The mean of~$\epsilon_-(\alpha)$ is zero because~$E[\nu(u)]=0$. From~Eq.~\ref{eq : covar_nu_BP}, we find that in the case of a photon noise:
\begin{equation}
\begin{array}{lrr}
\vspace{.2cm}
E[\epsilon_-(\alpha_1)\,\epsilon_-^*(\alpha_2)] =  \\
\displaystyle \int_{\mathrm{D}}\int_{\mathrm{D}}\frac{\mathcal{F}^{-1}[I](u_1-u_2)}{\lambda_0}\,\exp{\left[\frac{2\,i\,\pi\,(\alpha_2.u_2-\alpha_1.u_1)}{\lambda_0}\right]\,\mathrm{d}u_1\,\mathrm{d}u_2}
\end{array}
,\label{eq : covar_epsilon_-}
\end{equation}
where~$\displaystyle \int_{\mathrm{D}}$ represents the integral over a pupil of diameter~$D$. Calling~$\mathcal{A}_{\mathrm{D}}(\alpha)=J_1(\alpha)/\alpha$ where~$J_1$ is the Bessel function of the first kind corresponding to a~$\mathrm{D}$-diameter circular aperture, we can write
\begin{equation}
\begin{array}{lrr}
\vspace{.2cm}
E[\epsilon_-(\alpha_1)\,\epsilon_-^*(\alpha_2)] = \\
\displaystyle\iint{I(\alpha_3)\,\mathcal{A}_{\mathrm{D}}\left(\|\alpha_1-\alpha_3\|\right)\,\mathcal{A}_{\mathrm{D}}^*\left(\|\alpha_2-\alpha_3\|\right)\,\mathrm{d}\alpha_3}.
\end{array}
\label{eq : covar_epsilon_-bis}
\end{equation}
We deduce from~Eq.~\ref{eq : covar_epsilon_-bis} the variance of~$\epsilon_-$ in the case of photon noise:
\begin{equation}
Var[\epsilon_-(\alpha)]=Var[I_{-,n}(\alpha)]=I(\alpha)*\left|\mathcal{A}(\|\alpha\|)\right|^2.
\label{eq : var_epsilon_-_BP}
\end{equation}
Practically speaking, we record the interferential image~$I$ on a finite number of pixels and numerically process data. This means that the Fourier transform of~Eq.~\ref{eq : covar_epsilon_-} is a fast Fourier transform and the width of~$|\mathcal{A}|^2$ is~$\sim(\lambda_0/D)_{\mathrm{pix}}$\,pixels. Finally, the~$|\mathcal{A}|^2$ convolution in the expression of~$Var[\epsilon_-]$ roughly corresponds to averaging the noise over~$\sim(\lambda_0/D)^2_{\mathrm{pix}}$\,pixels and
\begin{equation}
Var[I_{-,n}(\alpha)]\simeq \frac{I(\alpha)}{(\lambda_0/D)^2_{\mathrm{pix}}}.
\label{eq : var_epsilon_-_B}
\end{equation}

A similar result is found for the noise on the unmodulated part~$I_{\mathrm{cent}}$ of the image~$I\simeq I_{\mathrm{S}}+I_{\mathrm{R}}+I_{\mathrm{C}}$~(Eq.~\ref{eq : icent_def}):
\begin{equation}
Var[I_{\mathrm{cent},n}(\alpha)]\simeq \frac{4\,I(\alpha)}{(\lambda_0/D)^2_{\mathrm{pix}}}.
\label{eq : var_icent_B}
\end{equation}
The sole difference is the size of the selecting binary mask in the correlation plane.



\end{document}